\documentclass[amsmath,amssymb,aps,longbibliography, superscriptaddress]{revtex4-2}
\numberwithin{equation}{section}

\usepackage{tikz}
\usetikzlibrary{quantikz2}
\usepackage[toc]{appendix}
\usepackage{tabularx}
\usepackage{soul}

\usepackage{bbm}
\usepackage{multirow}
\usepackage{float}
\usepackage{graphicx}
\usepackage{xcolor}
\usepackage{physics}
\usepackage{bbold}
\usepackage{chemformula}
\usepackage{siunitx}

\DeclareSIUnit\angstrom{\protect \text {Å}}

\usepackage[compat=0.6]{yquant}
\usetikzlibrary{fit,quotes}

\pgfdeclareshape{yquant-multictrlshape}{
   \inheritsavedanchors[from=yquant-circle]%
   \foreach \anc in {center, north, north east, east, south east, south, south west, west, north west} {%
      \inheritanchor[from=yquant-circle]{\anc}%
   }%
   \inheritanchorborder[from=yquant-circle]%
   \backgroundpath{%
      \pgfpathmoveto{\pgfqpoint{-0.707107\dimexpr\xradius\relax}{-0.707107\dimexpr\yradius\relax}}%
      \pgfpathlineto{\pgfqpoint{0.707107\dimexpr\xradius\relax}{0.707107\dimexpr\yradius\relax}}%
      \pgfpathellipse{\pgfpointorigin}%
                     {\pgfqpoint{\xradius}{0pt}}%
                     {\pgfqpoint{0pt}{\yradius}}%
   }%
   \inheritclippath[from=yquant-circle]%
}

\yquantset{multictrl/.style={every control/.style={shape=yquant-multictrlshape,radius=1.05mm}}}

\definecolor{riverlane_green}{RGB}{0, 150, 143}
\definecolor{riverlane_orange}{RGB}{255, 117, 0}

\usepackage{hyperref}
\hypersetup{
  colorlinks   = true, 
  urlcolor     = riverlane_green, 
  linkcolor    = riverlane_green, 
  citecolor   = riverlane_orange  
}
\usepackage[capitalise]{cleveref}

\newcommand{\oo}{\mathcal{O}}

\newcommand{\fes}{[\ch{Fe2S2}]$^{2-}$}

\newcommand{\br}{ {\bf r}}

\begin{document}
\title{Quantum Simulations of Chemistry in First Quantization with any Basis Set}
\author{Timothy N. Georges}
\affiliation{Riverlane Ltd., St Andrews House, 59 St Andrews Street, Cambridge, CB2 3BZ, United Kingdom}
\affiliation{Department of Chemistry, Physical and Theoretical Chemistry Laboratory, University of Oxford, Oxford, OX1 3QZ, United Kingdom}
\author{Marius Bothe}
\affiliation{Riverlane Ltd., St Andrews House, 59 St Andrews Street, Cambridge, CB2 3BZ, United Kingdom}
\affiliation{Astex Pharmaceuticals, 436 Cambridge Science Park, Cambridge, CB4 0QA, United Kingdom}
\author{Christoph S\"underhauf}
\affiliation{Riverlane Ltd., St Andrews House, 59 St Andrews Street, Cambridge, CB2 3BZ, United Kingdom}
\author{Bjorn K. Berntson}
\affiliation{Riverlane Ltd., St Andrews House, 59 St Andrews Street, Cambridge, CB2 3BZ, United Kingdom}
\author{R\'obert Izs\'ak}
\affiliation{Riverlane Ltd., St Andrews House, 59 St Andrews Street, Cambridge, CB2 3BZ, United Kingdom}
\author{Aleksei V. Ivanov}
\email[Corresponding author: ]{aleksei.ivanov@riverlane.com}
\affiliation{Riverlane Ltd., St Andrews House, 59 St Andrews Street, Cambridge, CB2 3BZ, United Kingdom}

\begin{abstract}
Quantum computation of the energy of molecules and materials is one of the most promising applications of fault-tolerant quantum computers. Practical applications require development of quantum algorithms with reduced resource requirements. Previous work has mainly focused on quantum algorithms where the Hamiltonian is represented in second quantization with compact basis sets while existing methods in first quantization are limited to a grid-based basis. In this work, we present a new method to solve the generic ground-state chemistry problem in first quantization using any basis set. We achieve asymptotic speedup in Toffoli count for molecular orbitals, and orders of magnitude improvement using dual plane waves as compared to the second quantization counterparts. In some instances, our approach provides similar or even lower resources compared to previous first quantization plane wave algorithms that, unlike our approach, avoids the loading of the classical data. The developed methodology can be applied to variety of applications, where the matrix elements of a first quantized Hamiltonian lack simple circuit representation.
\end{abstract}

\maketitle

\section{Introduction}

\label{sec:introduction}
Fault-tolerant quantum computing holds the potential to push the boundaries of quantum chemistry calculations~\cite{aspuru-guzikSimulatedQuantumComputation2005a,reiherElucidatingReactionMechanisms2017a,liElectronicComplexityGroundstate2019a,caoQuantumChemistryAge2019,bauerQuantumAlgorithmsQuantum2020a,liuProspectsQuantumComputing2022a,goingsReliablyAssessingElectronic2022c,bluntPerspectiveCurrentStateoftheArt2022a,delgado_simulating_2022,baiardiQuantumComputingMolecular2023,santagatiDrugDesignQuantum2024} due to efficient quantum algorithms, such as quantum phase estimation (QPE)~\cite{kitaevQuantumMeasurementsAbelian1995a} that, given a sufficient initial state preparation~\cite{leeEvaluatingEvidenceExponential2023}, allows for a nearly exact estimation of the energy of a molecule or a material in polynomial time. Specifically, QPE with qubitization is the leading approach \cite{lowHamiltonianSimulationQubitization2019c,poulin_quantum_2018,berryImprovedTechniquesPreparing2018a} which requires the lowest quantum resources for quantum chemistry problems~\cite{bluntPerspectiveCurrentStateoftheArt2022a,leeEvenMoreEfficient2021a,vonburgQuantumComputingEnhanced2021a,berryQubitizationArbitraryBasis2019a,reiher_elucidating_2017}. In order to use this algorithm, one must map the Hamiltonian and wavefunction into the qubit representation and also embed the Hamiltonian into the block of a larger unitary matrix using, for example, a linear-combination-of-unitaries (LCU) decomposition. The computational cost of the quantum algorithm is determined by the efficiency of these procedures, which in turn depends on the representation of the quantum chemistry Hamiltonian and whether it is in first or second quantization.

The most studied approach in the context of quantum computation is second quantization. In this formalism, the anti-symmetry of the electronic wavefunction is encoded into the creation and annihilation operators. These operators can be mapped into the qubit representation using, for example, the Jordan-Wigner transformation~\cite{Jordan1928,ortiz_quantum_2001}. Conveniently, the occupation number wavefunction maps directly onto the qubit basis, requiring $2D$ qubits for a system with $2D$ spin orbitals. The resulting computational cost does not explicitly depend on the number of electrons.
Many LCU decompositions of the second quantized Hamiltonian exist, including the sparse representation \cite{berryQubitizationArbitraryBasis2019a}, single factorization \cite{berryQubitizationArbitraryBasis2019a}, double factorization \cite{vonburgQuantumComputingEnhanced2021a,roccaReducingRuntimeFaulttolerant2024}, and tensor hyper-contraction \cite{leeEvenMoreEfficient2021a}. All of these methods, based on the factorization of the two-body term, are valid for any basis functions and can take advantage of state-of-the-art quantum chemistry basis sets.

There is, however, a tempting alternative. The first quantization formalism requires $N\log_2 2D$ qubits to represent the wavefunction, where $N$ is the number of electrons. As a result, for a fixed $N$, first quantization offers an exponential improvement in the scaling of the number of system qubits with respect to the number of orbitals~\cite{abramsSimulationManyBodyFermi1997a}. Therefore, we can increase the number of orbitals to obtain a better approximation of the continuum limit, with little increase in the number of system qubits~\cite{berryImprovedTechniquesPreparing2018a}. Furthermore, unlike methods developed in second quantization, the first quantization approach is equally applicable to bosonic problems with fixed number of particles as it is to fermionic ones; the only difference is the initial state preparation of the wavefunction which accounts for the symmetry.

Previous work on qubitization in first quantization is limited to plane-wave (PW) basis sets~\cite{babbushQuantumSimulationChemistry2019,suFaultTolerantQuantumSimulations2021} (for other approaches see Refs.~\cite{kassalPolynomialtimeQuantumAlgorithm2008a,tolouiQuantumAlgorithmsQuantum2013a,whitfield_communication_2013,whitfield_unified_2015,babbushExponentiallyMorePrecise2018b,chanGridbasedMethodsChemistry2023}). The electronic integrals have a simple analytical form in the PW basis set, which makes it possible to avoid the loading of classical data and instead use quantum arithmetic circuits to estimate the parameters of the Hamiltonian on the quantum computer. These methods have been developed further~\cite{ziniQuantumSimulationBattery2023a,berryQuantumSimulationRealistic2023a} to implement the Goedecker-Teter-Hutter pseudopotentials~\cite{goedecker_separable_1996,hartwigsen_relativistic_1998}. However, one of the main disadvantages of these approaches from an applications point of view is their inability to treat active regions of molecules or materials - one of the main tasks in computational quantum chemistry. While pseudopotentials do separate electrons into active (valence) and froze (core) electrons, this is not the same as a more general scheme for active space construction used in quantum chemistry. The other disadvantage is that it might not be possible to incorporate modern pseudopotentials~\cite{Hamann_1979_norm_conserving,Vanderbilt_soft_1990} or PAW method~\cite{blochl_projector_1994}, which do not have a simple analytical representation, without the classical-data loading. Therefore, developing approaches in first quantization for advanced quantum chemistry basis sets and exploring its advantages over second quantization in the context of QPE is of great interest. 

In this work, we present a qubitization-based QPE implementation of the quantum chemistry Hamiltonian in first quantization, with an arbitrary basis set. We provide the first explicit LCU decomposition in first quantization that makes use of a ``sparse" representation of the Hamiltonian. While we focus on electronic Hamiltonians, our result is general for any Hamiltonian in first quantization. We then explore two basis sets: molecular orbitals spanned on Gaussian-type orbitals and dual plane waves (DPW)~\cite{babbushLowDepthQuantumSimulation2018a}. For molecular orbitals, we compare quantum resource estimates of QPE with qubitization for active space calculations between sparse implementations in first and second quantization~\cite{berryQubitizationArbitraryBasis2019a}. One of the main quantum primitives in sparse qubitization is advanced quantum read only memory (QROAM~\cite{berryQubitizationArbitraryBasis2019a,low2018trading,babbushEncodingElectronicSpectra2018}) which allows for a trade-off between the number of qubits and Toffoli gates. Irrespective of the QROAM implementation, we surprisingly observe a polynomial speedup with respect to the number of basis functions, $D$, in the Toffoli count for a given number of electrons. For the implementation of QROAM that minimizes the number of logical qubits, the first quantization approach naturally provides exponential improvements in the qubit count, while for the implementation of QROAM that minimizes the number of Toffoli gates, we observe similar logical qubit count in both first and second quantization. The speedup in Toffoli-gate count is due to a lower subnormalization factor of the sparse LCU in first quantization. For DPWs, we also analyze the uniform electron gas and materials in the physically relevant regimes, and for DPWs, we observe a speed up of many orders of magnitude in both the logical qubit count as well as Toffoli counts over the second quantization counterpart. For some cases, we also observe resource reduction as compared to a first quantization plane wave (PW) implementation~\cite{suFaultTolerantQuantumSimulations2021} which avoids the data loading. To our best knowledge, this is the first work that implements first quantization in any basis sets for qubitized QPE.


\section{Results}
\subsection{LCU Decomposition in First Quantization}

\label{sec:lcu}
In first quantization, the Hamiltonian  of interacting identical particles can be written as follows
\begin{equation}\label{eq:full_ftq_chem_hamiltonian_with_spin}
    \hat{H} =
    \sum_{i=0}^{N-1}\sum_{p,q=0}^{D-1}\sum_{\sigma=0,1}h_{pq} \left(\ketbra{p\sigma}{q\sigma} \right)_{i}+
    \frac{1}{2}\sum_{i\ne j}^{N-1}\sum_{p,q,r,s=0}^{D-1}\sum_{\sigma,\tau=0,1} h_{pqrs}\left(\ketbra{p\sigma}{q\sigma}\right)_{i}  \left(\ketbra{r\tau}{s\tau}\right)_{j}.
\end{equation}
Here, $N$ is the number of particles, $D$ is the number of basis functions (orbitals) and $\sigma$ and $\tau$ are spin indices. The one-body and two-body matrix elements, $h_{pq}$ and $h_{pqrs}$, are independent of particle number and spin. The subscript on the operator $(\ketbra{p\sigma}{q\sigma})_i$ indicates that the operator acts on the $i^{\text{th}}$ particle:
\begin{equation}
    (\ketbra{p\sigma}{q\sigma})_i = I_{2D}^{\otimes i} \otimes \ketbra{p\sigma}{q\sigma} \otimes I_{2D}^{\otimes(N - i - 1)},
\end{equation}
where $I_{2D}$ is the identity operator acting on vector space ${\mathbb C}^{2D}$.

To block encode the Hamiltonian for QPE with qubitization, we can express this Hamiltonian as a linear combination of unitary matrices (LCU),

\begin{equation}\label{eq:LCU}\hat{H}_{\text{LCU}}=\sum_{\alpha}\omega_{\alpha}U_{\alpha},
\end{equation}
where $U_{\alpha}$ are unitary matrices with coefficients $\omega_{\alpha}$. 
The one-norm of the LCU,
\begin{equation}\label{eq:one-norm}\lambda=\sum_{\alpha}|\omega_{\alpha}|,
\end{equation}
is the subnormalisation of the LCU block encoding.

As a choice for unitary matrices we consider Pauli strings, which are tensor products of the Pauli matrices:
\begin{equation}
    I=\left(\begin{array}{cc}
        1 & 0 \\
        0 & 1
    \end{array}\right),\quad 
    X=\left(\begin{array}{cc}
        0 & 1 \\
        1 & 0
    \end{array}\right),\quad 
    Y=\left(\begin{array}{cc}
        0 & -\mathrm{i} \\
        \mathrm{i} & 0
    \end{array}\right),\quad 
    Z=\left(\begin{array}{cc}
        1 & 0 \\
        0 & -1
    \end{array}\right).
\end{equation}
To express the Hamiltonian as an LCU with Pauli strings, we use the generic Pauli LCU decomposition presented in Ref.~\cite{paulidecompriverlane2024}, which is valid for square matrices of dimension $D=2^{M}$ where $M$ is a natural number. The Pauli LCU decomposition of the one-body term then reads as 

\begin{equation}\label{eq:onebody_lcu_with_spin_conventional}
    \hat{H}_{\text{LCU},1} =
    \sum_{p,q=0}^{D-1} \omega_{pq} (-\mathrm{i})^{\mu (p, q)}
    \sum_{j=0}^{N-1}\prod_{k=0}^{M-1}X_{jM+k}^{p_k}Z_{jM+k}^{q_k}  \mathrm{i}^{p_k \wedge q_k} ,
\end{equation}
where $\mu (p, q) = \sum_{k=0}^{M-1} {p_k \wedge q_k}$, $\wedge$ is the AND operation, $X_a$ and $Z_a$ are Pauli operators acting on qubit $a$:
\begin{equation}    
\begin{split}    
    X_a & = I^{\otimes a} \otimes X \otimes I^{\otimes (MN - a - 1)}, \\
    Z_a & = I^{\otimes a} \otimes Z \otimes I^{\otimes (MN - a - 1)},  
\end{split}
\quad a = 0 \dots MN - 1
\end{equation}
and $p_k$ is the value of the $k^{\text{th}}$ bit of the standard binary representation of $p$, likewise for $q_k$. The action on the spin index is the identity, so there are $N\log D$ system qubits. The coefficients $\omega_{pq}$ of the Pauli strings are given by an XOR transform of the indices followed by a Hadamard transform,
\begin{equation}\label{eq:onebody_omega_def}
    \omega_{pq} = \frac{1}{D} \sum_{a=0}^{D-1} h_{p\oplus a,a} \big(H^{\otimes M}\big)_{a, q},
\end{equation}
where $\oplus$ is the bitwise XOR operation and $H$ is a Hadamard matrix, defined as~\footnote{In quantum computation Hadamard matrix is defined with a prefactor of $\frac{1}{\sqrt{2}}$ unlike our definition}:
\begin{equation}~\label{eq:hadamard}
    H = 
\begin{pmatrix}
    1 & 1 \\
    1 & -1
\end{pmatrix}
\end{equation}
While the Eq.~\eqref{eq:onebody_lcu_with_spin_conventional} is written using the conventional definition of Pauli strings, we find it more convenient to cancel $-\mathrm i$ with $\mathrm i$ and work with a slightly different form:
\begin{equation}\label{eq:onebody_lcu_with_spin}
    \hat{H}_{\text{LCU},1} =
    \sum_{p,q=0}^{D-1} \omega_{pq}
    \sum_{i=0}^{N-1}\prod_{k=0}^{M-1}X_{Mi+k}^{p_k}Z_{Mi+k}^{q_k}.
\end{equation}
For a real Hamiltonian, anti-Hermitian terms in Eq.~\eqref{eq:onebody_lcu_with_spin} have  coefficients $\omega_{pq}$ equal to zero, and all remaining unitary matrices are also Hermitian. From now on, we will refer to this form of the decomposition as Pauli LCU.
The two-body LCU decomposition is
\begin{equation}\label{eq:twobody_lcu_with_spin}
    \hat{H}_{\text{LCU},2} =
    \sum_{p,q,r,s=0}^{D-1} \omega_{pqrs}
    \frac{1}{2}\sum_{i\ne j}^{N-1}
    \prod_{k=0}^{M-1}X_{Mi+k}^{p_k}Z_{Mi+k}^{q_k} X_{Mj+k}^{r_k}Z_{Mj+k}^{s_k},
\end{equation}

\noindent where the two-body Pauli LCU coefficients $\omega_{pqrs}$, are

\begin{equation}\label{eq:twobody_omega_def}
    \omega_{pqrs} = \frac{1}{D^2}\sum_{a=0}^{D-1}\sum_{b=0}^{D-1}
    h_{p\oplus a,a,r\oplus b,b}\big(H^{\otimes M}\big)_{a,q} \big(H^{\otimes M}\big)_{b, s} .
\end{equation}
\cref{eq:twobody_lcu_with_spin} is a straightforward generalization of Ref.~\cite{paulidecompriverlane2024} to a 4-dimensional tensor and details of this generalization are given in~\cref{apx:derivation_lcu_twobody}. \cref{eq:onebody_omega_def,eq:twobody_omega_def} are manipulated into a matrix multiplication so that the fast Walsh-Hadamard transform can be used to efficiently find the LCU coefficients~\cite{paulidecompriverlane2024}. 

Coefficients that correspond to the same Pauli string can be combined to reduce the one-norm of the LCU and data loading cost. When $p=q=0$ or when $r=s=0$, the two-body LCU repeats Pauli strings from the one-body LCU. Moreover, all terms that give the identity may be removed; these terms shift the eigenvalues of the Hamiltonian by a constant and do not need to be input on the quantum computer. These changes give the canonicalized one-body LCU,

\begin{equation}\label{eq:canon_lcu_h1}\hat{H}_{\text{LCU},1}^{\prime}=\sum_{p,q=0}^{D-1}
    \omega^{\prime}_{pq}
    \sum_{i=0}^{N-1}\prod_{k=0}^{M-1}X_{Mi+k}^{p_k}Z_{Mi+k}^{q_k},
\end{equation}

\noindent where

\begin{equation}
    \omega^{\prime}_{pq} = \begin{cases}
        0, & \text{if } p=q=0 \\
    \omega_{pq}+\frac{1}{2}\left(N-1\right)
    \left(\omega_{pq00}+\omega_{00pq}\right), & \text{otherwise.}
    \end{cases}
\end{equation}

\noindent The canonicalized two-body LCU is

\begin{equation}\label{eq:canon_lcu_h2}
    \hat{H}_{\text{LCU},2}^{\prime} =
    \sum_{p,q,r,s=0}^{D-1} \omega^{\prime}_{pqrs}
    \frac{1}{2}\sum_{i\ne j}^{N-1}
    \prod_{k=0}^{M-1}
    X_{Mi+k}^{p_k}Z_{Mi+k}^{q_k}
    X_{Mj+k}^{r_k}Z_{Mj+k}^{s_k},
\end{equation}

\noindent where

\begin{equation}\label{eq:canonical_coeff_2b}
    \omega^{\prime}_{pqrs} = \begin{cases}
        0, & \text{if } p=q=0 \text{ or } r=s=0 \\
    \omega_{pqrs}, & \text{otherwise.}
    \end{cases}
\end{equation}

\noindent The removed terms proportional to the identity are
\begin{equation}\label{eq:canon_lcu_h0}
    \hat{H}_{\text{LCU},0}^{\prime} = N\omega_{00} + \frac{1}{2}N(N-1)\omega_{0000}.
\end{equation}
The cost of an LCU block encoding depends on the number of unique, non-zero coefficients, $L$, and the one-norm, $\lambda$. The upper bound of the number of non-zero terms from \cref{eq:onebody_lcu_with_spin,eq:twobody_lcu_with_spin} is $L=D^4+D^2$. The one-norm of the canonicalized LCU is

\begin{equation}\label{eq:ftq_one-norm}
    \lambda = \sum_{p,q=0}^{D-1}|\omega^{\prime}_{pq}|\cdot N +
    \sum_{p,q,r,s=0}^{D-1}|\omega^{\prime}_{pqrs}| \cdot \frac{1}{2}N(N-1).
\end{equation}
\noindent 
 In the case of dense matrix elements $h_{pqrs}$, such that $\sum_{pqrs} |h_{pqrs}| = O(D^4) $, we expect $\lambda=O(N^2D^3)$. This is because in ~\cref{eq:twobody_omega_def}, $H^{\otimes M}/\sqrt{D}$ is the unitary transformation, and there is an extra prefactor $1/D$. We confirm this speedup numerically in ~\cref{subsec:scaling_properties_mol} for dense matrix elements. However, for chemical applications, $h_{pqrs}$ is sparse and $\lambda$ depends on the particular basis set and symmetries presented in the system. Our numerical simulations (\cref{subsec:scaling_properties_mol,subsec:scaling_properties_dpw}) show that for molecular orbitals we observe less than a $1/D$ speedup, while for dual plane waves we achieve more than $1/{D}$ compared to second quantization for a fixed $N$. If $D=O(N)$, the one-norm in second quantization scales better than in first quantization in the examples we studied. However, for a fixed $N$, $\lambda$ can be smaller than the one-norm in second quantization not only in the asymptotic regime, but also for a practical size of the basis set. We present detailed numerical analyzes for different basis sets in \cref{subsec:scaling_properties_mol,subsec:scaling_properties_dpw} and show that for dual plane waves this is indeed the case.

So far, our results are completely general for any Hamiltonian. We used no symmetries of the tensors $h_{pq}$ and $h_{pqrs}$, or of the coefficients $\omega_{pq}$ and $\omega_{pqrs}$, to derive \cref{eq:onebody_lcu_with_spin} through \cref{eq:ftq_one-norm}. 
In ~\cref{sec:symmetries}, we discuss the consequences of different symmetries and the choice of the basis functions on the LCU.

\subsection{Matrix Elements and Symmetries in Different Basis Sets}
\label{sec:symmetries}
\subsubsection{Arbitrary Basis Sets}\label{sec:sym_chem_basis_sets}

In quantum chemistry, the two-body matrix elements of the Hamiltonian can be computed as

\begin{equation}
    h_{pqrs} = \iint_{V\times V} d\br d\br'  \frac{\phi^{*}_{p}(\br) \phi_{q}(\br) \phi^{*}_{r}(\br') \phi_{s}(\br')}{|\br - \br'|} ,
\end{equation}

\noindent where $\phi_{q}(\br)$ are molecular orbitals and $V$ is the supercell in the case of a periodic solid, or is  ${\mathbb R}^3$ for an isolated molecule. Often, these orbitals are chosen to be real and in this case the two-body matrix elements possess the $8$-fold symmetry

\begin{equation}\label{eq:8fold_symmetry}
    h_{pqrs} = h_{qprs} = h_{pqsr} = h_{qpsr} = 
    h_{rspq} = h_{srpq} =  h_{rsqp} =  h_{srqp}   .
\end{equation}
\noindent One-body matrix elements are symmetric for real orbitals,
\begin{equation}\label{eq:2fold_symmetry}
    h_{pq} = h_{qp}.
\end{equation}

Data loading is often a dominant cost of qubitized QPE. To reduce this cost, symmetries can be exploited to load only the unique, non-zero LCU coefficients. Next, we discuss how the symmetries of real matrix elements, given in \cref{eq:8fold_symmetry,eq:2fold_symmetry}, affect the one- and two-body LCU coefficients in turn.

The one-body LCU coefficients do not retain the two-fold symmetry of the one-body matrix elements. However, the Hadamard and XOR transforms of the matrix elements convert the symmetry into zero coefficients. The number of non-zero matrix elements for the one-body term is less than or equal to $\frac{1}{2}D(D+1)$~\cite{paulidecompriverlane2024}. 
By examining \cref{eq:twobody_lcu_with_spin}, it can be seen that the two-body LCU coefficients retain a two-fold symmetry,
\begin{equation}\label{eq:twobody_2fold_symmetry}
    \omega_{pqrs}=\omega_{rspq}.
\end{equation}
This symmetry simplifies the canonicalized one-body coefficients so that
\begin{equation}
    \omega^{\prime}_{pq} = \begin{cases}
        0, & \text{if } p=q=0 \\
    \omega_{pq}+\left(N-1\right)
    \omega_{pq00}, & \text{otherwise.}
    \end{cases}
\end{equation}
Accounting for the XOR and Hadamard transform of the LCU, the remaining two-fold symmetry and canonicalisation, the number of unique, non-zero, two-body LCU coefficients is less than or equal to $D(D+1)(D-1)(D+2)/8$. Indeed, assuming all original tensor elements are non-zeros, for each pair $(p, q)$ that corresponds to non-zero value of $\omega_{pqrs}$ we have $K:=D(D+1)/2$ pairs of $(r, s)$. This corresponds to $K^2$ non-zero coefficients. Taking into account the symmetry of $\omega_{pqrs}$ under $(p, q)\leftrightarrow (r, s)$ and canonicalization~\eqref{eq:canonical_coeff_2b}, we obtained the total number of unique, non-zero coefficients to be $(K^2-K)/2=D(D+1)(D-1)(D+2)/8$.

\subsubsection{Basis Sets Diagonalizing the Coulomb Potential}
We next consider the Hamiltonian in basis sets which diagonalize the Coulomb potential, such as dual plane waves. Then,
\begin{equation}
\label{eq:diagonal_hamiltonian}
    \hat H = \sum_{p,q=0}^{D-1}T_{pq}\sum_{i=0}^{N-1}\left(\ketbra{p}{q}\right)_{i}
    + 
    \frac{1}{2}
    \sum_{p,r=0}^{D-1}V_{pr}
    \sum_{i\ne j}^{N-1}
    \left(\ketbra{p}\right)_{i} \left(\ketbra{r}\right)_{j},
\end{equation}where $T_{pq}$ and $V_{pr}$ are symmetric matrices. This Hamiltonian transforms to a Pauli LCU as:
\begin{equation}\label{eq:HV_PWD_qubit}
    \hat{H} = 
    \sum_{p,q=0}^{D-1}
    \omega_{pq}
    \sum_{i=0}^{N-1}\prod_{k=0}^{M-1}
    X_{Mi+k}^{p_k}Z_{Mi+k}^{q_k}
    +
    \sum_{p,r=0}^{D-1}
    \gamma_{pr}\cdot 
    \frac{1}{2}\sum_{i\ne j}^{N-1}\prod_{k=0}^{M-1}
    Z_{Mi+k}^{p_k} Z_{Mj+k}^{r_k},
\end{equation}
with the LCU coefficients
\begin{equation}\label{eq:dpw_plcu_coeff_one_body}
    \omega_{pq} = \frac{1}{D}\sum_{u=0}^{D-1}
    T_{p\oplus u,u} \big(H^{\otimes M}\big)_{uq}
\end{equation}
and
\begin{equation}\label{eq:dpw_plcu_coeff_two_body}
    \gamma_{pr} = \frac{1}{D^2}\sum_{u,v=0}^{D-1} \big(H^{\otimes M}\big)_{pu} V_{uv} \big(H^{\otimes M}\big)_{vr}.
\end{equation}
As before, symmetries of the original one-body matrix elements manifest in the sparsity of $\omega_{pq}$, while two-body coefficients $\gamma_{pr}$ remain symmetric.
Similarly to the previous case, there is double-counting of Pauli strings in one- and two-body terms. By removing contributions proportional to the identity, the canonicalized Hamiltonian in the Pauli basis can be written as follows
\begin{equation}\label{eq:HV_PWD_canonical_qubit}
    \hat{H}_1 + \hat{H}_2 = 
    \sum_{p,q=0}^{D-1}
    \omega^{\prime}_{pq}
    \sum_{i=0}^{N-1}\prod_{k=0}^{M-1}
    X_{Mi+k}^{p_k}Z_{Mi+k}^{q_k}
    +
    \sum_{p,r=1}^{D-1}
    \gamma'_{pr}\cdot 
    \frac{1}{2}\sum_{i\ne j}^{N-1}\prod_{k=0}^{M-1}
    Z_{Mi+k}^{p_k} Z_{Mj+k}^{r_k},
\end{equation}
where the one-body and two-body matrix elements are now 
\begin{equation}
    \omega^{\prime}_{pq} = \begin{cases}
        \omega_{pq}, & \text{if } p > 0 \\
    \omega_{0q} + (N-1)\gamma_{0q}, & \text{if } p=0 \text{ and } q>0 \\
    0, & \text{if } q=0 \text{ and } p=0.
    \end{cases}
\end{equation}
\begin{equation}
    \gamma'_{pr} =  \begin{cases}
        0, & \text{if } p \text{ or } r=0 \\
        \gamma_{pr} & \text{otherwise.}
    \end{cases}
\end{equation}
Here we again disregard terms with identity Pauli-strings, since they only contribute a constant shift of the eigenvalue.
An example of such basis sets are DPWs~\cite{babbushLowDepthQuantumSimulation2018a}, which exhibit additional translation symmetry. This symmetry manifests in the sparsity of Pauli LCU coefficients as will be numerically shown in \cref{subsec:scaling_properties_dpw}.

\subsection{Block Encoding}

\label{sec:block_encoding}

In qubitized QPE, the Hamiltonian is embedded into a block of the larger unitary matrix $U$ with subnormalization factor $\lambda$:
\begin{equation}
    \hat U = \begin{pmatrix}\hat H/\lambda & \ast \\  \ast &  \ast \end{pmatrix}.
\end{equation}
This unitary is then used to construct the quantum-walk operator,
\begin{equation}
    \hat W = \hat U\begin{pmatrix}I & \\ & -I\end{pmatrix},
\end{equation}
which has eigenvalues related to the spectrum of the Hamiltonian, ${\rm Eigenvalue}[\hat W]_i = e^{\pm \arccos\left(E_i/\lambda\right)}$~\cite{poulin_quantum_2018,berryImprovedTechniquesPreparing2018a}.

In a circuit implementation of the block encoding of $\hat{U}$, the value of the junk blocks is irrelevant. 
We will use a linear combination of unitaries (LCU) approach to constructing the block encoding.
A block encoding can be constructed for an LCU $\hat H = \sum_{l=0}^{L-1} |a_l| \hat U_l$ by combining the operators PREP (state preparation) and SELECT:
\begin{equation}
    \text{PREP}\ket{0}^{\otimes\log_2 L} = \sum_{l=0}^{L-1}\sqrt{|a_l|}\ket{l},\ \text{SELECT} = \sum_{l=0}^{L-1} \ket{l}\!\bra{l}\otimes \hat U_l,\ \lambda =\sum_{l=0}^{L-1} |a_l|,\ \hat U = \text{PREP}^\dagger\cdot\text{SELECT}\cdot\text{PREP}.
\end{equation}
Typically, the dominant cost is $O(\sqrt{L}\lambda/\epsilon_{\rm QPE})$, which assumes a QROAM-based implementation of PREP \cite{berryQubitizationArbitraryBasis2019a}.

First, we will show the circuit constructions for the sparse qubitization in first quantization for arbitrary basis sets, based on the Hamiltonian \eqref{eq:canon_lcu_h1} and \eqref{eq:canon_lcu_h2}, and then, we discuss how the circuits cost reduces for the Hamiltonian with a diagonal Coulomb interaction~\eqref{eq:HV_PWD_canonical_qubit}.

\subsubsection{Arbitrary Basis Sets}\label{subsec:arbitrarybasisset_blockencoding}
Let us  write both the one-body \eqref{eq:canon_lcu_h1} and the two-body \eqref{eq:canon_lcu_h2} Hamiltonian as a single sum of $i\neq j$. For the one-body term, we use $s=r=0$:
\begin{align}
    \hat H' = \hat H'_\text{LCU,1} + \hat H'_\text{LCU,2} = \sum_{pqrs=0}^{D-1} a_{pqrs} \sum_{i\neq j}^{N-1}
     \prod_{k=0}^{M-1} X^{p_k}_{iM+k}
     Z^{q_k}_{iM+k}
    \prod_{k'=0}^{M-1} X^{r_{k'}}_{jM+k'}
     Z^{s_{k'}}_{jM+k'}
     \label{eq:H arbitrary}
\end{align}
with coefficients
\begin{equation}
        a_{pqrs} = 
    \begin{cases}
        \omega'_{pq}/(N-1) & \text{for}\ r=s=0 \\
        \omega'_{pqrs}/2 & \text{for}\ (p,q) = (r,s) \\
        \omega'_{pqrs} & \text{for}\ (p,q)<(r,s) \\
        0 & \text{otherwise}.
    \end{cases}
    \label{eq:arbitrary coefs}
\end{equation}
The inequality $(p,q)<(r,s)$ should be understood as inequality of composite indices and selects one branch of the symmetry $(p,q)\leftrightarrow(r,s)$. The factor of $1/(N-1)$ in the one-body coefficients cancels the extra sum over $j$.

{\it A priori}, there are $L=D^4$ coefficients $a_{pqrs}$ to load. However, the coefficients are very sparse with many zeros or values close to zero that can be truncated, resulting in an effective $L<D^4$, which can be harnessed by using ideas from the sparse qubitization scheme~\cite{berryQubitizationArbitraryBasis2019a}. Therefore, we will load $L$ coefficients $a_l$, along with index value $p(l),q(l),r(l),s(l)$ for use in SELECT.

\textit{The SELECT operator.}
We begin by showing a circuit construction for a SELECT operator selecting $\prod_{k=0}^{M-1}Z^{q_k}_{iM+k}$. We denote the circuit as SELECT${}_{i,q}[\prod_{k=0}^{M-1}Z^{q_k}_{iM+k}]$ where the indices are the control registers for the SELECT operation.
Our approach is to iterate over $i$ with a unary iteration gadget \cite{babbushEncodingElectronicSpectra2018}. For each value of $i$, we apply $M=\log_2 D$ CCZ gates, each of which targets the $iM+k$'th system qubit ($k=0,\ldots, M-1$) and is controlled on the unary iterator qubit specifying the current value of $i$ as well as the $k$'th qubit of the $q$ register. This is illustrated in the following, where we only show the unary iterator qubit and hide all other ancillary qubits and inner workings of the unary iteration. Between the first two cnots, the unary iterator qubit is $\ket{1}$ if $i=0$. Between the second two cnots, the unary iterator qubit is $\ket{1}$ if $i=1$, etc.
\begin{equation}
\scalebox{0.8}{%
\begin{tikzpicture}
\begin{yquant}
qubit control;

qubit {$\ket{i}$} i;

qubit {unary iterator} iter;
discard iter;

qubit {} p[4];
init {$\ket{q}$} (p);
discard p;
text {$\vdots$} p[2];
init {$\ket{q_0}$} p[0];
init {$\ket{q_1}$} p[1];
init {$\ket{q_{M-1}}$} p[3];

qubit {system $\ket{\psi}$} system;

["north:$NM$" {font=\protect\footnotesize, inner sep=0pt}]
slash system;

["north:$ \lceil \log_2 N \rceil$" {font=\protect\footnotesize, inner sep=0pt}]
slash i;


[name=first]
cnot i | control;

cnot iter | i;

settype {qubit} iter;
box {$Z_{0M + 0}$} system | iter, p[0];
box {$Z_{0M + 1}$} system | iter, p[1];
text {$\cdots$} iter, p, system;
box {$Z_{0M + (M-1)}$} system | iter, p[3];
cnot iter | i;
box {$Z_{1M + 0}$} system | iter, p[0];
box {$Z_{1M + 1}$} system | iter, p[1];
text {$\cdots$} iter, p, system;
[name=ireg]
box {U} i;
box {$Z_{1M + (M-1)}$} system | iter, p[3];
cnot iter | i;
text {$\cdots$} iter, p, system;
[name=last]
cnot iter | i;
discard iter;

\end{yquant}
\node[draw, fill=white, fit=(first-0) (ireg-0) (last-p0)] {unary iteration};
\end{tikzpicture}
}
\label{eq:basic select}
\end{equation}

The above circuit can be adjusted to implement $\text{SELECT}_{i,p,q}\left[\prod_{k=0}^{M-1}X^{p_k}_{iM+k}Z^{q_k}_{iM+k}\right]$. Rather than implementing it as a product of two circuits \eqref{eq:basic select} for $q$ and $p$, we combine them to save the cost of a second unary iteration: A CCX gate is added to the left of each CCZ gate in \eqref{eq:basic select}, and is controlled on the relevant qubit of the $\ket{p}$ register instead of the $\ket{q}$ register. This has cost
    \begin{equation}
\text{SELECT}_{i,p,q}\left[
    \prod_{k=0}^{M-1}X^{p_k}_{iM+k}Z^{q_k}_{iM+k}\right]:\ (N-1)+2NM\ \text{Toffolis},\ \lceil \log_2 N \rceil  \ \text{ancilla qubits},
\end{equation}
when using measurement based uncomputation for the unary iteration. The $(N-1)$ Toffolis and ancilla qubits are internal to the unary iteration. 

For the LCU \eqref{eq:H arbitrary}, we use the circuit twice to implement the product
\begin{equation}    \text{SELECT}_{i,p,q}\left[
    \prod_{k=0}^{M-1}X^{p_k}_{iM+k}Z^{q_k}_{iM+k}\right]\cdot\text{SELECT}_{j,r,s}\left[\prod_{k=0}^{M-1}X^{r_k}_{jM+k}Z^{s_k}_{jM+k}\right].
    \label{eq:arbitrary full select}
\end{equation}
The same circuit can be used for both one-body and two-body LCU terms, by setting the control registers $r=s=0$ in the one-body terms, as is taken care of in \eqref{eq:arbitrary coefs}.

The work of \cite{suFaultTolerantQuantumSimulations2021} uses a different approach to SELECT: Rather than implementing gates for each value of $i$ in turn \eqref{eq:basic select}, the part of the system register belonging to electron $i$ is swapped towards the top, then the gates are implemented on that register once, and it is swapped back. In our case this is not advantageous as we are not carrying out arithmetic on the system register and the CCX and CCZ gates are cheaper than the swaps and extra unary iteration for the unswaps would be.

\textit{The PREPARE operator.}
As mentioned, we use a sparse scheme to reduce the data loading cost in the face of many zero coefficients or small coefficients that are truncated to zero. For this, the non-zero coefficients $a_{p,q,r,s}$ from the LCU~\eqref{eq:H arbitrary} are indexed by $l\in\{0,\ldots L-1\}$ in arbitrary order.
We use the sparse scheme to construct the PREP operator with QROAM and coherent alias sampling, see \cite{berryQubitizationArbitraryBasis2019a,leeEvenMoreEfficient2021a,ivanovQuantumComputationPeriodic2023}. The construction of the PREP circuit is largely unchanged as compared to sparse qubitization in second quantization and the main modification is an additional step for preparing the equal superposition over electron numbers. For completeness, we provide a detailed description below:
\begin{enumerate}
    \item Prepare an equal superposition state $\frac{1}{\sqrt L}\sum_{l=0}^{L-1}\ket{l}$ of dimension $L$ via amplitude amplification (see \cref{apx:amplification}). 
    \item Data lookup: Use a QROAM indexed on $\ket{l}$ to load indices and sign $\ket{p,q,r,s,\theta}$, alternative indices and sign, and keep probabilities for the coefficients $a_{pqrs}$. This is the dominant cost of PREP in both time and space.
    \item Coherent alias sampling: Perform swaps between indices and alternative indices based on an inequality test between keep probability and an equal superposition state. The sign qubit does not have to be swapped but the phase can be applied directly (omitted during uncomputation).
    \item Prepare an equal superposition state $\frac{1}{\sqrt{N(N-1)}}\sum_{i\neq j}^{N-1}\ket{i}\!\ket{j}$. This can be done by preparing an equal superposition state on $2\lceil \log_2 N\rceil$ qubits and doing amplitude amplification, see \cref{apx:amplification}.
\end{enumerate}
The SELECT operator from \eqref{eq:arbitrary full select} should be controlled on the successful outcome of the uniform superposition preparations in step 1 and 4, requiring an extra Toffoli gate. 
Uncomputation of PREP can make extensive use of measurement based uncomputation, such that the only Toffoli cost is that of uncomputing the QROAM and the equal superposition states.
The walk operator required for QPE is the product of the block encoding and a reflection. The reflection must be on the flag qubits, which are $\lceil\log_2L\rceil + 2$ for the equal superposition state $\frac{1}{\sqrt{L}}\sum_{l=0}^{L-1}\ket{l}$ and its success qubit and rotation qubit for amplitude amplification, similarly $2\lceil\log_2N\rceil + 2$ for $\frac{1}{\sqrt{N(N-1)}}\sum_{i\neq j}^{N-1}\ket{i}\!\ket{j}$. The costs are summarised in \cref{table:costs}.

\begin{table}
\begin{tabular}{|l|l|l|}\hline
Circuit element & Toffoli count & qubits \\\hline\hline
\textbf{walk operator:} & & \\
system register & n.a. & $NM$ \\
PREP equal superposition $\frac{1}{\sqrt L}\sum_{l=0}^{L-1}\ket{l}$ & $3\lceil \log_2 L\rceil - 3\eta_L + 2 b_{L} -9$ & $b_L+2$\\
PREP equal superposition $\frac{1}{\sqrt{N(N-1)}}\sum_{i\neq j}^{N-1}\ket{i}\!\ket{j}$ & $ 8 \lceil \log_2 N \rceil - 4\eta_N+2 b_N -7$& $b_N+2$ \\
PREP data lookup via QROAM & $\lceil L/\kappa_1\rceil + m(\kappa_1-1)$ & $m\kappa_1 + \lceil\log(L/\kappa_1)\rceil$ \\
PREP coherent alias sampling & $\aleph + (m-\aleph-2)/2$ & 0 \\
SELECT (for arbitrary basis) & $2(N - 1 + 2NM + 1)$& 0\\
SELECT (for basis diagonalising Coulomb) & $2N + 3NM$&0 \\
UNPREP coherent alias sampling & 0 & 0 \\
UNPREP data lookup via QROAM & $\lceil L/\kappa_2\rceil + \kappa_2$ & 0 \\
UNPREP equal superpositions & like PREP steps& 0 \\
reflection & $\lceil\log_2L\rceil + 2\lceil\log_2N\rceil +2$ & 0
\\\hline
\textbf{additional QPE cost per step:} & &\\
phase estimation ancillas & n.a. & $\lceil\log_2(\mathcal{I}+1)\rceil$ \\
unary iteration over walk operator & 1 & $\lceil\log_2(\mathcal{I}+1)\rceil - 1$ \\
make reflection controlled & 1 & 0
\\\hline
\textbf{total} & $(\sum \text{above})\cdot \mathcal{I}$ & $(\sum \text{above})$ \\\hline
\end{tabular}

\vspace*{3mm}

\begin{tabular}{ll}
Variable & description \\\hline
$N$ & number of electrons \\
$D$ & number of basis functions (orbitals) \\
$M = \log_2 D$ & number of qubits per electron (apart from spin)\\
$L$ & number of non-zero coefficients in LCU, see \eqref{eq:arbitrary coefs}\\
$\eta_L$,$\eta_N$ & largest power of 2 that is a factor of $L$ or $N$\\
$b_L,b_N = 8$ & precision of rotation for equal superposition preparations \\
$\kappa_1$ & power of 2, space-time trade-off parameter for QROAM; chosen to minimise Toffoli cost \\
$\kappa_2$ & power of 2, space-time trade-off parameter for QROAM uncomputation; chosen to minimise Toffoli cost \\
$m = \aleph + 2(4M + 1)$ & output size of QROAM (for arbitrary basis)\\
$m = \aleph + 2(3M + 1)$ & output size of QROAM (for basis diagonalising Coulomb) \\
$\aleph$ & bits for keep-probabilities in coherent alias sampling \\
$\mathcal I = \lceil \pi\lambda/(2\epsilon_\text{QPE})\rceil$  & number of repetitions of walk operator for QPE
\end{tabular}
\caption{{\bf Cost for one walk operator and the total QPE circuit.} As usual in qubitization literature, we neglect the one-off costs that are not multiplied by $\mathcal{I}$ (initial state preparation for the system register, initial state preparation for the QPE ancillas, inverse QFT for the QPE and the preparation of the catalytic state used in amplitude amplification (see \cref{apx:amplification})).
Furthermore, we assume that the space-time tradeoff for the QROAM is such that sufficient ancilla qubits are available to avoid additional ancilla qubits in later steps. Extensive use of measurement-based uncomputation reduces the Toffoli cost of UNPREP.}
\label{table:costs}
\end{table}

\subsubsection{Basis Sets Diagonalizing the Coulomb Potential} \label{sec:diagonal_coloumb}
The Hamiltonian~\eqref{eq:HV_PWD_canonical_qubit} for block-encoding can be rewritten as
\begin{equation}
    \hat H' = \sum_{pqr=0}^{D-1} a_{pqr} \sum_{i\neq j}^{N-1}
     \prod_{k=0}^{M-1}
     \left(
     X^{p_k}_{iM+k}
     Z^{q_k}_{iM+k} \right)
    \prod_{k'=0}^{M-1}
     Z^{r_{k'}}_{jM+k'},
     \label{eq:H pwdual}
\end{equation}
where the combined matrix elements are 
\begin{equation}
    a_{pqr} = \begin{cases}
        \omega^{\prime}_{pq}/(N-1), & \text{if } r=0 \\
    \gamma_{qr}, & \text{if } p=0 \text{ and } 0<q<r \\
    \gamma_{qr}/2, & \text{if } p=0 \text{ and } q=r\ne 0 \\
    0, & \text{otherwise.}
    \end{cases}
\end{equation}
Compared to the costs for a general Hamiltonian shown in \cref{table:costs} resource requirements differ in several ways: The number of unique coefficients $L$ is much smaller, less than or equal to $D^2 + D(D+1)/2$, and we never need to implement an $XZXZ$-type Pauli string, so the cost of the second SELECT operator does not have the controlled $X$-gates, which lowers the Toffoli cost to $2N + 3 NM$, as opposed to $2N + 4 NM$ Toffolis. In addition we only need to specify 3 indices in the output of the QROAM register in the PREP step, allowing for $m=\aleph + 2(3M+1)$ instead of $m=\aleph + 2(4M+1)$. These Toffoli counts are also indicated in Table~\ref{table:costs}.

\subsection{Scaling Properties of Pauli LCU and Quantum Resource Estimates}
\label{sec:qpe_costing}
In this section, we first discuss the properties of the LCU decomposition and establish asymptotic scalings which we compare to existing methods. The properties that affect the overall scaling of QPE are the one-norm and the number of unique, non-zero coefficients. A good LCU minimises both of these factors. We will calculate the one-norm and number of unique, non-zero coefficients for several systems, truncating either the matrix elements of the Hamiltonian or its LCU coefficients. 

With these results, we then provide the quantum resource estimates of QPE with qubitization for a range of systems. We focus on single-shot QPE resource estimation and do not consider subdominant costs such as the antisymmetrization of the initial state~\cite{berryImprovedTechniquesPreparing2018a}. To obtain the number of physical qubits, we will assume that the algorithm is run on superconducting qubits placed on a square grid. We consider the code cycle duration of $10^{-6}$ s and the physical error rate of $0.1 \%$ -- conventional figures for superconducting qubits~\cite{krinner_realizing_2022,acharya_suppressing_2022}. We employ the surface code following Refs.~\cite{litinski_magic_2019,litinski_game_2019,Gidney2019a,Campbell2017a,Jones2013a,Eastin2013,fowler_surface_2012}; for details we refer the reader to Appendix D of Ref.~\cite{ivanov_upaw_2024}.

Implementation of qubitization-based QPE uses QROAM which allows for a trade-off between the number of logical qubits and the number of Toffoli gates. Therefore, we will consider two versions of the algorithm: one minimises the number of logical qubits by not using additional copies of the input registers for data loading, which corresponds to setting $\kappa_1 = 1$ in \cref{table:costs} (min-Qu). The other version of the algorithm chooses $\kappa_1$ such that it minimises the number of Toffoli gates needed instead (min-T).

The systems that we will consider are a dense, real Hamiltonian, the square-planar \ch{H4} molecule, a hydrogen chain, the \fes complex, the uniform electron gas and the crystalline solids nickel oxide and diamond. We apply the DPW basis set to the periodic systems and for the others we use molecular orbitals. These systems range from test models to possible applications. 

We compare our work primarily against the sparse, second quantization LCU of Ref.~\cite{berryQubitizationArbitraryBasis2019a} in \cref{subsec:scaling_properties_mol,subsec:quantum_resources_mol}. We aim to establish what advantage first quantization brings over second quantization when matrix elements of the Hamiltonian are treated similarly without applying any additional factorization. We believe that this method is comparable to our method as a baseline from which other methods in the same quantization can be developed. Additionally, we will compare our results with the second quantization DPW method of Ref.~\cite{babbushLowDepthQuantumSimulation2018a} and the PW approach of Ref.~\cite{babbushQuantumSimulationChemistry2019,suFaultTolerantQuantumSimulations2021} in \cref{subsec:resources_dpw}.

\subsubsection{Molecular Orbitals. Scaling Properties for Basic Models.}
\label{subsec:scaling_properties_mol}
\begin{figure}
    \centering
    \includegraphics[width=18cm]{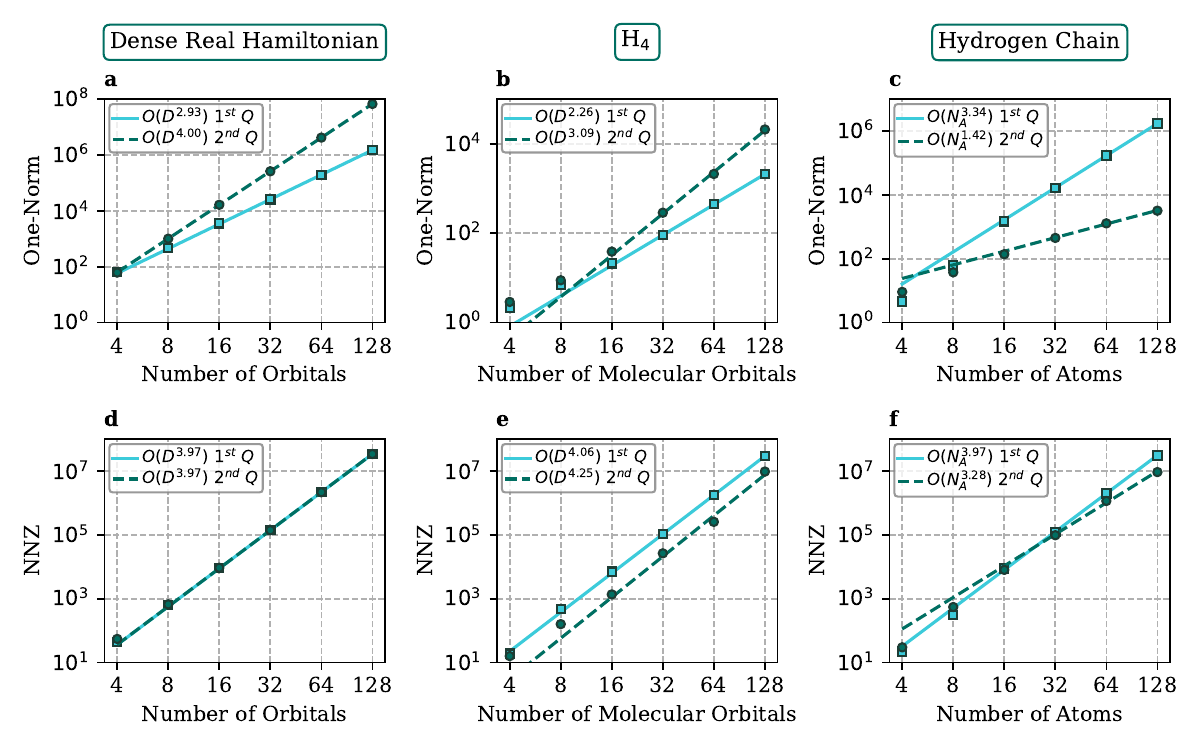}
    \caption{\textbf{Scaling of the one-norm and number of unique, non-zero coefficients (NNZ) for test systems.} \textbf{a} For a dense, real Hamiltonian with a fixed number of electrons, the one-norm of first quantization scales as $\mathcal{O}(D^3)$ compared to $\mathcal{O}(D^4)$ for the equivalent second quantization method, where $D$ is the number of orbitals. \textbf{b} Similarly, for \ch{H4} the one-norm scales better in first than in second quantization. \textbf{c} Hydrogen chains with varying number of atoms, $N_A$, with the number of orbitals per atom, $N_A/D$, being constant. The number of electrons, $N$, is proportional to the number of atoms and orbitals, $N = O(N_A)=O(D)$. Therefore, the $\mathcal{O}(N^2)$ dependence of the one-norm in first quantization leads to worse scaling. \textbf{d} For a dense, real Hamiltonian, the number of unique, non-zero terms is the same for first and second quantization. \textbf{e} For the \ch{H4} molecule, the number of unique non-zero terms is greater for first quantization. \textbf{f} The number of unique, non-zero terms has better scaling in second quantization than first for the Hydrogen chain.
    }

    
    \label{fig:bigplot1}
\end{figure}

In order to benchmark this work against Berry's approach~\cite{berryQubitizationArbitraryBasis2019a}, our first point of comparison is generating random matrix elements to form a dense, real Hamiltonian. The one-body matrix elements have two-fold symmetry and the two-body matrix elements have eight-fold symmetry. This is a worst case scenario of scaling for a generic Hamiltonian. For this reason, we do not truncate these matrix elements nor the LCU coefficients, except for a $10^{-10}$ Ha cut-off for `zero' coefficients. To compare the asymptotic scaling of these methods, only the dominant two-body term is considered. For a dense real Hamiltonian, we keep the number of electrons constant at $N=4$, so that it does not affect the scaling in $D$. The one-norm scales as $\oo(D^{2.93})$ in first quantization and as $\oo(D^{4})$ in second quantization as shown in \cref{fig:bigplot1}(a). \cref{fig:bigplot1}(d) shows that both methods have $\oo(D^4)$ unique, non-zero LCU coefficients as expected. 

Square planar \ch{H4} is a common molecule used to compare different algorithms. In this work, we aim to use the same bond lengths and truncation procedures as Lee et al.~\cite{leeEvenMoreEfficient2021a}, so that we can compare our scaling results. The \ch{H4} molecule has $N=4$ electrons and a bond length of $2$ Bohr radii. For the second quantization representation, we truncate the matrix elements so that there is an error of less than $5.0\times 10^{-5}$ Ha per hydrogen atom, using CCSD(T). We use the aug-cc-pVQZ~\cite{dunning_gaussian_1989,kendall_electron_1992} basis with $D=4$ to $128$ orbitals, using Boys localization. For the first quantization representation, we truncate the LCU coefficients to achieve this error instead, because the Pauli LCU decomposition is reversible and it is the number of non-zero LCU coefficients that determine the computational cost of the algorithm. As before, we count all LCU coefficients below $10^{-10}$ as zero and we only consider the two-body term since we are interested in the asymptotic scaling of the methods. For the line of best fit, we use the final three data points. For \ch{H4}, the one-norm has lower asymptotic scaling in first quantization. As shown in \cref{fig:bigplot1}(b), our method scales as $\oo(D^{2.26})$, while the Berry representation scales as $\oo(D^{3.09})$. The Berry representation has fewer unique, non-zero LCU coefficients, but a slightly worse asymptotic scaling of $\oo(D^{4.25})$ compared to $\oo(D^{4.06})$ is obtained, see Fig.~\ref{fig:bigplot1}(e). However, we expect that the asymptotic scaling is actually $\oo(D^4)$, which is obscured because the truncation improved some data points more than others.

The hydrogen chain is another common system used to compare methods. It was also considered in Lee et al.~\cite{leeEvenMoreEfficient2021a} and we will similarly use a bond length of $1.4$ Bohr radii and a truncation procedure that aims for a CCSD(T) error of $5.0\times 10^{-5}$ Ha per hydrogen atom employing STO-3G basis set. In such basis set, we have one orbital per atom, so $N=D=N_A$, where $N_A$ is the number of atoms in the chain. As for \ch{H4}, we truncate the matrix elements of the second quantization representation and the LCU coefficients of the first quantization representation. Unlike the previous two examples, the number of electrons increases with the number of orbitals for the hydrogen chain model. For the previous systems, the number of system qubits is smaller in first quantization. For the hydrogen chain, the dependence of the number of orbitals on the number of electrons turns the system qubit number advantage into a disadvantage. In this case, first quantization requires $D\log D$ system qubits compared to $2D$ in second quantization. Because the number of electrons grows with the system size, the $\oo(N^2)$ scaling of the first quantization one-norm, shown in \cref{eq:ftq_one-norm}, becomes an $\oo(D^2)$ scaling. Therefore, the scaling of the one-norm in first quantization for this problem is about $D^2$ worse than second quantization methods, as shown in \cref{fig:bigplot1}(c). There is no such dependence on the number of electrons in the number of non-zero LCU coefficients. However, the scaling of the Berry representation is superior at $\oo(D^{3.28})$, compared to $\oo(D^{3.97})$ for the first quantization representation, see Fig.~\ref{fig:bigplot1}(f).

\subsubsection{Quantum Resource Estimates for Iron-Sulfur Complex}
\label{subsec:quantum_resources_mol}

Iron-sulphur clusters are known for their structural versatility and are ubiquitous in biochemical systems \cite{beinert1997iron,rouault2017iron}. They easily participate in redox reactions in which electrons are exchanged between reactants and as a result form a crucial part of electron transfer mechanisms that take place in several important enzymes. These clusters consist of a Fe$_n$S$_m$ ($n,m$ are some integers) core cluster to which a number of ligands are attached and in an enzyme several such clusters may form an electron transfer pathway to ensure the passage of electrons within the protein \cite{volbeda1995crystal,pandelia2013electronic}. There are several other functions iron-sulfur clusters serve in biological systems, and they also represent a challenge to electronic structure theory calculations \cite{beinert1997iron,rouault2017iron}. The characterization of magnetic interactions in these clusters is especially challenging. In a typical scenario, each iron in the cluster can be thought of as a site where several unpaired electrons with parallel spins are localized. The electromagnetic properties of the entire cluster then depend on the interactions between several of these sites. Antiferromagnetic coupling is often favoured between two sites, in which case all electrons in one site are aligned in an antiparallel fashion compared to the electrons on the other site. In chemistry, this situation perhaps is one of the best candidates for finding strong correlation effects~\cite{izsak2023measuring,ganoe2024notion}. Consequently, such systems require demanding classical computational techniques for even a qualitatively accurate calculation, and even the simplest of these clusters, such as Fe$_2$S$_2$, often feature as test systems in classical computational studies \cite{sharmaLowenergySpectrumIron2014a,dobrautz2021spin}. Such clusters are also used as a benchmark molecule in the context of quantum computation~\cite{tazhigulov2022simulating,leeEvaluatingEvidenceExponential2023}.

In this work, we compare the quantum resource estimates of our new approach with the second quantization sparse qubitization algorithm of Berry \textit{et al.}~\cite{berryQubitizationArbitraryBasis2019a}, considering the \fes complex as a test system. While the active space we consider consists of only 14 electrons, this is a more realistic example than those we used in \cref{subsec:scaling_properties_mol}. In order to define the active space Hamiltonian, we have followed methodology similar to Refs.~\cite{leeEvaluatingEvidenceExponential2023,sharmaLowenergySpectrumIron2014a}. Namely,
we used the cc-pVTZ-DK basis set~\cite{balabanov_systematically_2005, de_parallel_2001, woon_gaussian_1993, dunning_gaussian_1989}
and carried out high-spin restricted open-shell Kohn-Sham calculations with the exact two-component (x2c) scalar-relativistic correction~\cite{liu_ideas_2010,saue_relativistic_2011,peng_exact_2012} and the BP86 functional~\cite{becke_density-functional_1988,perdew_density-functional_1986}. We achieve convergence with the second-order co-iterative augmented Hessian (CIAH) method~\cite{sun_co_iterative_2016}. 
Next, we localised the virtual and occupied spaces separately using Pipek-Mezey (PM) localisation \cite{lehtola_pipek_2014, pipek1989fast}. We choose $12$ occupied orbitals for the active space, consisting of five $3d$ orbitals per iron atom and one $3p_z$ orbital per bridging sulfur atom. The active space is then completed by selecting $D-12$ orbitals from the virtual space based on their orbital energy such that $D=16, 32, 64, 128$. The first $122$ virtual space orbitals are all centred on iron or a bridging sulfur.

In order to carry out resource estimations, one has to distribute the error budget among all approximations used in the quantum algorithm~\cite{leeEvenMoreEfficient2021a}:
\begin{equation}
    \epsilon_{\rm tot} = \epsilon_{\rm QPE} + \epsilon_{\rm trunc} + \epsilon_{\rm PREP},
\end{equation}
where $\epsilon_{\rm QPE}$ is the error due to finite precision in QPE, $\epsilon_{\rm trunc}$ is the error due to truncation of the Hamiltonian and $\epsilon_{\rm PREP}$ is the error in the state preparation (PREP circuit). We choose these parameters as follows~\cite{leeEvenMoreEfficient2021a}:
\begin{equation}
    \epsilon_{\rm QPE} = \frac{10}{16} \epsilon_{\rm tot}, 
\quad \epsilon_{\rm trunc} = \frac{3}{16} \epsilon_{\rm tot}, \quad \epsilon_{\rm PREP} = \frac{3}{16} \epsilon_{\rm tot}.
\end{equation}
 Our target is to estimate the ground state energy with precision of $\epsilon_{\rm tot}=0.1$ mHa, since the energy difference between different spin states in \fes is on the order of only few mHa (the singlet-triplet gap is $\approx$ 2.0 mHa)~\cite{sharmaLowenergySpectrumIron2014a}. This is higher precision than that typically used in resource estimation for quantum chemistry (1.6 mHa).  We truncate the first quantization LCU coefficients and second quantization Hamiltonian matrix elements to meet the target error, $\epsilon_{\rm trunc}$, as estimated with an unrestricted MP2 (UMP2) calculation. We choose UMP2 method instead of CCSD(T) because UMP2 requires fewer computational resources and does not have problems with convergence, unlike CCSD(T) for large active spaces (128 orbitals).

\begin{figure}[H]
    \centering
    \includegraphics[width=18cm]{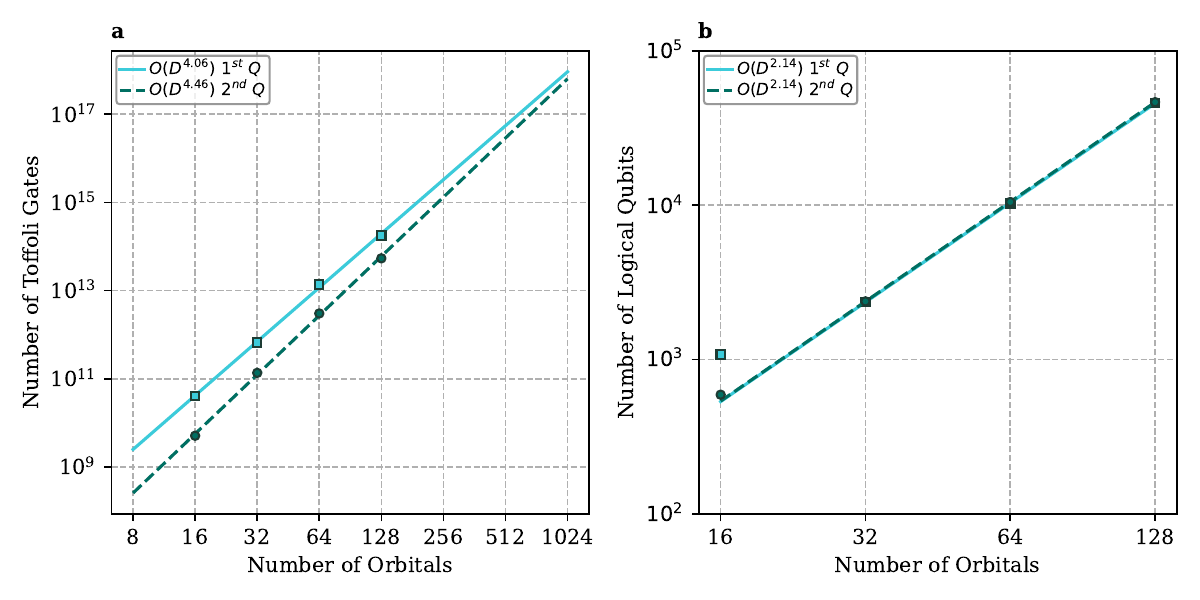}
    \caption{\textbf{Resource estimates of sparse qubitization in first and second quantization for \fes with 14 electrons, varying the number of orbitals.} Resource estimates were obtained for the target accuracy $\epsilon_{\rm tot}=0.1$ mHa for ground state energy estimation. \textbf{a} Second quantization requires fewer Toffoli gates for all data points, but first quantization scales better in $D$. \textbf{b} Only the last three points were used for the fit of logical qubits. First and second quantization require a similar number of logical qubits.
    }
    \label{fig:fe2s2_toffoli_qubits_01mha}
\end{figure}

\cref{fig:fe2s2_toffoli_qubits_01mha} shows the resource estimates for the two approaches, which both use QROAM that minimizes the Toffoli count (min-T). The number of Toffoli gates in first quantization scales asymptotically better by a factor of $D^{0.4}$. This asymptotic speedup is due to better scaling of the Pauli LCU one-norm in first quantization. The number of logical qubits is similar in both methods because of the small truncation value, $\epsilon_{\rm trunc}$. For $D\leq 1024$, sparse qubitization in second quantization shows smaller Toffoli count. The reason is that the one-norm of the first-quantization LCU explicitly depends on the number of electrons squared and this gives a large prefactor. Only at more than $10^{3}$ orbitals would we expect first quantization to show an advantage. 

\cref{tab:minimizing_qubits} shows the resource estimates when one chooses to minimize the number of logical qubits. As one can see, the number of qubits becomes smaller in first quantization compared to second quantization for more than 16 orbitals. For example, for 128 orbitals the number of qubits is 293 while in second quantization this number is 492. With larger system size there will be even larger improvement. However, the Toffoli count is again only asymptotically better in first quantization, scaling with order $\oo(D^{6.04})$ against $\oo(D^{6.23})$.

\begin{table}[H]
    \centering
    \begin{tabularx}{0.6\textwidth} { 
   >{\centering\arraybackslash}X 
   >{\centering\arraybackslash}X 
   >{\centering\arraybackslash}X  
   >{\centering\arraybackslash}X 
   >{\centering\arraybackslash}X   
  }
         \hline
          Orbitals & \multicolumn{2}{c}{Toffoli Gates} & \multicolumn{2}{c}{Logical Qubits} \\             
                  & 1\textsuperscript{st} & 2\textsuperscript{nd} 
                  & 1\textsuperscript{st} & 2\textsuperscript{nd}  \\
        \hline
16 & $1.79 \cdot 10^{11}$ & $2.16 \cdot 10^{10}$ & 198 & {\bf 190} \\
32 & $1.14 \cdot 10^{13}$ & $2.04 \cdot 10^{12}$ & {\bf 229} & 250 \\
64 & $9.52 \cdot 10^{14}$ & $1.44 \cdot 10^{14}$ & {\bf 261} & 340 \\
128 & $4.76 \cdot 10^{16}$ & $9.29 \cdot 10^{15}$ & {\bf 293} & 492 \\
        \hline
    \end{tabularx}
    \caption{\textbf{Resource estimates with QROAM that minimizes the number of qubits instead of the number of Toffolis.} Estimates are obtained for the total error budget of 0.1 mHa. The number of Toffolis is always smaller in second quantization. Bold font indicates the smallest number of qubits.}
    \label{tab:minimizing_qubits}
\end{table}
\subsubsection{Scaling properties using Dual Plane Waves} \label{subsec:scaling_properties_dpw}
In order to take advantage of the more efficient algorithm for systems with diagonal Coloumb interactions outlined in \cref{sec:diagonal_coloumb} we consider the DPW. This basis set is related to PW via a Fourier transform, so the basis functions are localised in real space around evenly spaced grid points (see Refs.\cite{skylaris2002nonorthogonal,mostofi2002total} and Appendix C of Ref.~\cite{babbushLowDepthQuantumSimulation2018a} for details). This transformation diagonalizes the Coloumb interaction, bringing the Hamiltonian into the form of \cref{eq:diagonal_hamiltonian} and greatly reducing the number of coefficients that need to be loaded. The trade-off is that one typically needs far more DPWs to achieve the same basis set accuracy compared to Gaussian orbitals, but since the size of our system register only scales logarithmically with the basis size we expect DPW to be favourable for our algorithm. 


The Pauli-LCU one-norm of the Hamiltonian~\eqref{eq:diagonal_hamiltonian} is bound by

\begin{equation}
     \oo\left(N D^{1/2}\right) \leq \lambda \leq  \oo\left(\frac{N(N+1)D^2}{2}\right) 
\end{equation}

\begin{figure}[H]
    \centering
    \includegraphics[width=18cm]{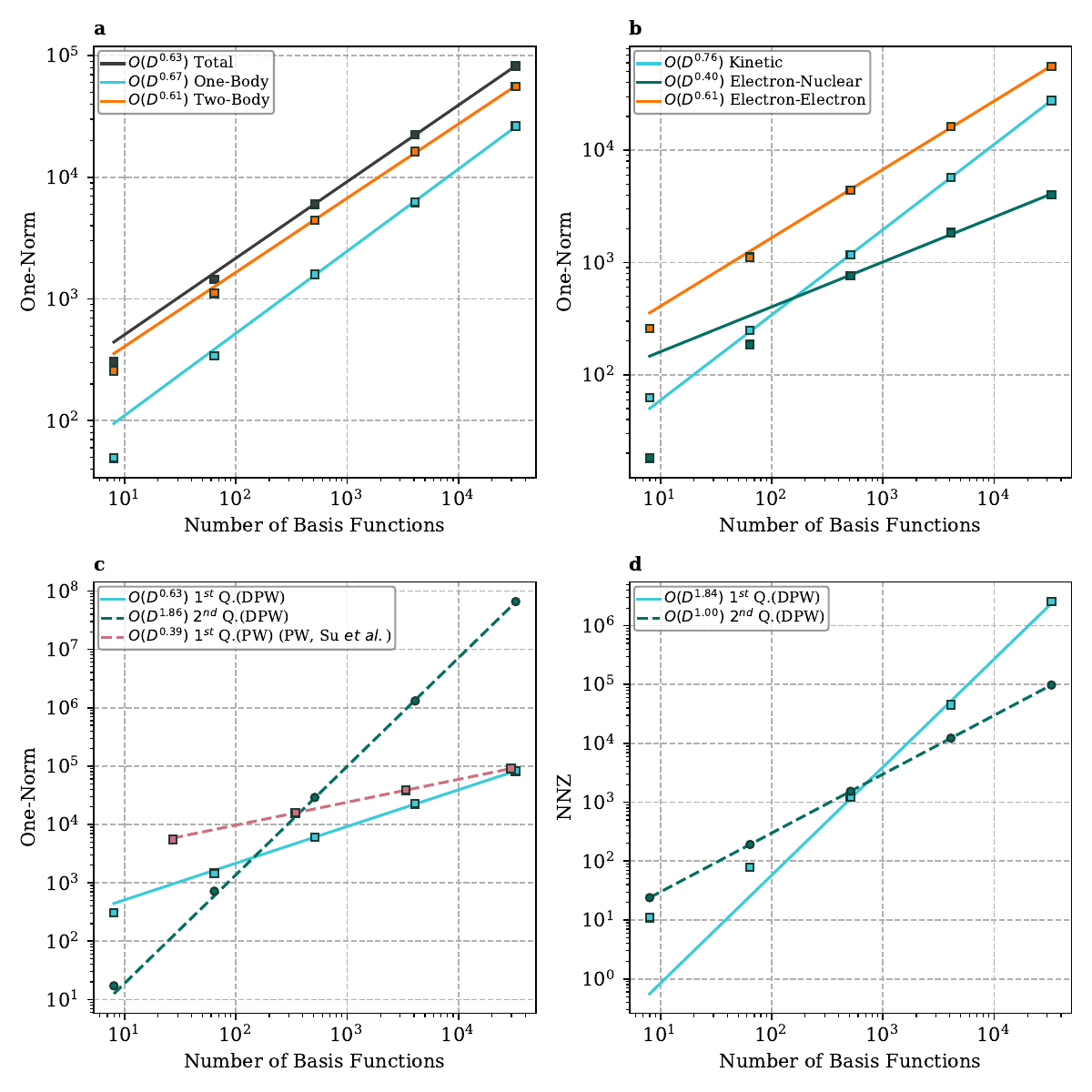}
    \caption{\textbf{Scaling of the one-norm and number of unique, non-zero coefficients (NNZ) for diamond.} Diamond is represented by an 8-atom unit cell \textbf{a} First quantization using DPWs. The two-body term contributes more to the one-norm. \textbf{b} First quantization using DPWs. Electron-electron interactions contribute more to the one-norm than the electron-nuclear or kinetic terms. \textbf{c} In first quantization, the scaling of the one-norm with respect to the number of basis functions, $D$, is more favourable than second quantization. The norm of PW approach of Ref.~\cite{suFaultTolerantQuantumSimulations2021} has the most favorable scaling but in the pre-asymptotic regime the first quantization DPWs norm of this work is lower.
    \textbf{d} The number of unique, non-zero coefficients scales as $\mathcal{O}(D^{1.84})$ in the number of basis functions, $D$, for first quantization using DPWs and as $\mathcal{O}(D)$ for second quantization using DPWs. 
    }
    \label{fig:diamond_norm_nnz}
\end{figure}

%
%
We note that this bound might be too loose for DPW and in order to establish a more precise dependence on the number of basis functions, we have carried out numerical calculations. For this, we have considered diamond with a cubic cell made of 8 atoms. The one-norm of the Pauli LCU consists of one- and two-body contributions and it is bounded by the sum of the one-norms of the kinetic energy, nuclear-electron and electron-electron interactions
\begin{equation}~\label{eq:}
    \lambda = \lambda_1 + \lambda_2 \leq \lambda_T + \lambda_U + \lambda_V.
\end{equation}
The total norm, as well as each individual contribution, is shown in \cref{fig:diamond_norm_nnz}(a,b). The one-body term provides smaller contribution to the total norm than the two-body term. The nuclear-electron contribution is around an order of magnitude smaller than electron-electron contribution and demonstrates subdominant scaling in basis functions. Based on these numerical results, we can conclude that the total norm scales as
\begin{equation}
    \lambda = \oo\left(N \left(\frac{D}{V}\right)^{2/3} D^{0.1}\right) + 
    \oo\left(N^2 \left(\frac{D}{V}\right)^{1/3} D^{0.28}\right),
\end{equation}
where the first contribution is due to the kinetic energy and the second term is due to the electron-electron interaction.
We note that the calculations we have done might be in the pre-asymptotic regime and as a result this scaling might change for the larger basis sets. However, this is the regime which is of interest when such methodology will be used with pseudo-potentials. In comparison, the one-norm in second quantization with DPW~\cite{babbushLowDepthQuantumSimulation2018a} scales as $\oo(D^{2})$ with respect to the number of DPWs and thus, we achieved a considerable speedup by using the same basis but employing first quantization. For a very small number of basis functions, the norm in second quantization is smaller than our norm but for the physically interesting regime we achieve orders of magnitude improvement as shown in \cref{fig:diamond_norm_nnz}{\bf c}. 

Apart from improvement in the one-norm, we also observe polynomial speedup in the space-time cost of the walk operator. Such cost consists of the SELECT and PREP circuits. In second quantization, the dominant cost will be SELECT in which the Toffoli-gate count scales linearly with the size of the basis set, $D$,~\cite{babbushEncodingElectronicSpectra2018} while in our approach such cost scales logarithmically in $D$. The most dominant cost is then due to data loading in PREP, where both Toffoli and qubit numbers scale as $\oo(\sqrt{\Gamma})$, where $\Gamma$ is the amount of information needed to specify the Hamiltonian~\cite{berryQubitizationArbitraryBasis2019a}. From \cref{fig:diamond_norm_nnz}{\bf d}, we can see that $\Gamma = \oo(D^{1.85})$ and as a result the Toffoli complexity scales sublinearly. We also note that we did not use any translation symmetries here and we expect that if one encorporates those, one can achieve $\oo(D^{1/2})$ scaling of PREP. When one uses QROM instead of QROAM to minimize the number of qubits, the space cost scales logarithmically in $D$, while the time cost of the block-encoding scales as $\oo(D^{1.85})$. Even in this case, we achieve smaller total resources, as will be shown in the next section.
%
%

\subsubsection{Resource Estimates for Materials}\label{subsec:resources_dpw}
In this section, we report quantum resource estimates for the uniform electron gas (UEG), diamond and nickel oxide (NiO). 
We compare our approach with the second quantization DPW algorithm~\cite{babbushEncodingElectronicSpectra2018} and the first quantization PW algorithm described in~\cite{suFaultTolerantQuantumSimulations2021}. The latter bypasses the data loading and uses quantum arithmetic to approximate the Hamiltonian matrix elements on the quantum computer.
The DPW can be obtained by applying the unitary transformation to PW and as a result, the same size DPW and PW would provide the same accuracy and can therefore be compared directly. However, the algorithm from Ref.~\cite{suFaultTolerantQuantumSimulations2021} can only work with basis sets of size $(2^n-1)^3$ for integer $n$, whereas our algorithm requires $2^{3n}$ basis states, and as a result, we cannot compare the identical basis sizes and instead use the closest matches. Since we do not introduce the error due to truncation of the Hamiltonian matrix elements, we distribute the total error, $\epsilon_{\rm tot}$, as $\epsilon_{\rm QPE} = 15.8 \epsilon_{\rm tot}/16$ and $\epsilon_{\rm PREP} = 0.2 \epsilon_{\rm tot}/16 $. The rationale for this is to reduce the Toffoli count as much as possible which is mainly affected by $\epsilon_{\rm QPE}$. 

We provide resource estimates for the UEG with $N=14, 54, 114$ electrons in the classically challenging strongly correlated regime. In order to converge the total energy of 14 electrons with 1 mHa accuracy, one would need on the order of 2000 plane waves, and the strongly correlated regime appears when the Winger-Seitz radius, $r_s$, is larger than 3.5 Bohr~\cite{lee_auxiliary-field_2019}. As shown in Ref.~\cite{lee_auxiliary-field_2019}, even for around 100 basis functions there is a significant deviation (a few tens of mHa) of approximate methods from i-FCIQMC~\cite{shepherd_investigation_2012}. The energy estimates in the regime with more than 1000 PWs and 14 electrons are not currently available with FCIQMC method due to high computational cost, although we do not state that this is impossible to reach with modern large-scale HPC. A larger system with $N=54$, and $r_s\ge 5$ is also challenging regime for classical algorithms, where different flavours of quantum Monte-Carlo algorithms give energies differing by more than 1 mHa per electron. For $N=114$, only few data points are available in the regime $r_s \leq 2$, and $D \leq 2109$~\cite{kwon1998effects,lee_auxiliary-field_2019}. Therefore, we consider $r_s=5$ and the number of basis functions, $D$, to be 4096. For the 14-electron system, we also present results for $D=512$. Resource estimates for these systems are shown in \cref{tab:UEG_costing}. Comparison between the four algorithms show that our method substantially outperforms the previous DPW algorithm by Babbush \textit{et al.}~\cite{babbushEncodingElectronicSpectra2018}, requiring only a fraction of logical qubits and reducing the Toffoli count by four orders of magnitude. The min-Qu algorithm also achieves the lowest logical qubit counts across all tested systems, by requiring a much smaller system register compared to the second quantization algorithm, whose system register scales linearly with the number of basis states. The number of logical qubits in the PW algorithm also scales logarithmically in $D$ similarly to the min-Q algorithm, but the qubit overhead of the PW algorithm is larger, and it requires more logical qubits overall. This results into higher physical qubit requirements, despite having overall lower quantum volume for large basis sets.

Next, we consider two examples of realistic materials: diamond and nickel oxide (NiO) in cubic cells with 8 and 64 atoms. While diamond is a relatively simple material to describe using classical electronic structure methods, NiO is the quintessential system for studying the strong correlations due to presence of localized $d$-electrons and it is often used as a prototype system to test the accuracy of different electronic structure methodologies~\cite{cui_efficient_2020,zhang_auxiliary-field_2018, peng_synergy_2017, mitra_many-body_2015, korotin_construction_2008}. All-electron calculations of such a material would require enormous PW basis sets. For example, the much simpler Si would require on the order of $10^{9}$ PW per \AA$^3$ to converge the mean-field eigenvalues with the accuracy of 0.01 eV using the \textit{regularized} Coulomb potential~\cite{gygi_all-electron_2023}. Instead of analyzing such a large basis set, we will focus on the practical size of basis sets which are suitable~\cite{lejaeghere_reproducibility_2016,prandini2018precision} for use with pseudopotentials~\cite{Hamann_1979_norm_conserving,Vanderbilt_soft_1990} or the projector augmented-wave method~\cite{blochl_projector_1994}. For example, a simple conversion from grid-spacing, $h$, to plane-wave cutoff $E_{\rm cut} = \frac{1}{2}(\pi / h)^2$~\cite{Briggs1996_real_space}, for 64-atom diamond and NiO cells would provide $E_{cut} \approx 750$ eV and $E_{cut} \approx 550$ eV kinetic energy cutoff, respectively. 
Resource estimates with the parameters described are shown in the \cref{tab:materials_costing}. As one can see, the min-Qu algorithm provides the lowest number of logical qubits for small cell calculations and as a result the lowest number of physical qubits, apart from the case of diamond and nickel oxide with 64 atoms cell, where the PW algorithm has lower physical qubit count. However, the number of physical qubits in both approaches differ by less than 10\%. The min-T algorithm provides the lowest Toffoli count in all cases apart from diamond represented with a 64-atom cell, where the first quantization PW approach is better. The interesting result is that for a large system with 1152 electrons and $\approx 32000$ basis functions, the min-T algorithm which loads the data from QROAM requires only a factor of 1.8 more logical qubits as compared to the first quantization PW approach which completely avoids the data loading. Similarly to the results obtained for the UEG model, the first quantization approach with DPWs (both min-Qu and min-T versions) is significantly better than the second quantization approach regardless of quadratic dependence of the first quantized Hamiltonian on the number of electrons as indicated in Table~\ref{tab:materials_costing}, and allows calculation of realistic systems with lower physical qubit counts than any prior method.

In order to understand why our approach provides a lower Toffoli count for some systems compared to the PW algorithm of Ref.~\cite{suFaultTolerantQuantumSimulations2021}, we report the Toffoli cost for construction of the quantum walk operator and the one-norm.  The one-norm of our approach is lower up to a few tens of thousands of basis functions (\cref{fig:diamond_norm_nnz}\textbf{d}.). 
In~\cref{tab:block_encoding_break_down}, we show the cost of block-encoding and the one-norm for each system. For diamond and nickel-oxide the norm of our approach is lower than that of Ref.~\cite{suFaultTolerantQuantumSimulations2021}, and for NiO with 8 atoms in the unit cell, the cost of block encoding is also slightly lower. For NiO with 64 atoms, the block encoding cost is slightly larger by a factor of 1.2, but this cost is compensated by a lower norm which leads to overall smaller Toffoli cost. For diamond with a 64 atom cell, the norm is smaller by a factor of 1.3 but the block encoding cost is larger by a factor of 2.0, which makes the total Toffoli cost larger. For larger basis sets (more then 30 000 basis functions) and for systems studied here, we expect the PW approach of Ref.~\cite{suFaultTolerantQuantumSimulations2021} to be better due to lower asymptotic scaling.

\begin{table}[H]
    \centering
    \begin{tabularx}{1.0\textwidth}{
   >{\centering\arraybackslash}c |
   >{\centering\arraybackslash}X |
   >{\centering\arraybackslash}X | 
   >{\centering\arraybackslash}X |
   >{\centering\arraybackslash}X | 
   >{\centering\arraybackslash}X
  }
         \hline
          System & Method & (Electrons, Orbitals) & Logical qubits & Toffolis & Physical qubits\\
\hline
UEG-14 & This work (min Qu) & (14, 512) & \bf 259& $2.01 \cdot 10^{7}$  & $\bf 6.83 \cdot 10^{5}$  \\
  & This work (min T) &   & 482& $\bf 1.32 \cdot 10^{7}$  & $1.32 \cdot 10^{6}$  \\
  & $2^{nd}$ Q. (DPW)\cite{babbushEncodingElectronicSpectra2018} &   & 1127& $2.96 \cdot 10^{10}$  & $4.87 \cdot 10^{6}$  \\
  & $1^{st}$ Q. (PW)\cite{suFaultTolerantQuantumSimulations2021} & (14, 343) & 593& $2.27 \cdot 10^{7}$  & $1.6 \cdot 10^{6}$  \\ \hline
UEG-14 & This work (min Qu) & (14, 4096) & \bf 330& $1.99 \cdot 10^{9}$  & $\bf 1.23 \cdot 10^{6}$  \\
  & This work (min T) &   & 1751& $2.36 \cdot 10^{8}$  & $5.71 \cdot 10^{6}$  \\
  & $2^{nd}$ Q. (DPW)\cite{babbushEncodingElectronicSpectra2018} &   & 8327& $1.27 \cdot 10^{13}$  & $5.69 \cdot 10^{7}$  \\
  & $1^{st}$ Q. (PW)\cite{suFaultTolerantQuantumSimulations2021} & (14, 3375) & 767& $\bf 8 \cdot 10^{7}$  & $2.39 \cdot 10^{6}$  \\
\hline

UEG-54 & This work (min-Qu) & (54, 4096) & \bf 813& $4.21 \cdot 10^{9}$  & $\bf 3.11 \cdot 10^{6}$  \\
  & This work (min-T) &   & 2249& $6.22 \cdot 10^{8}$  & $7.83 \cdot 10^{6}$  \\
  & $2^{nd}$ Q. (DPW)\cite{babbushEncodingElectronicSpectra2018}  &   & 8319& $1.93 \cdot 10^{12}$  & $4.88 \cdot 10^{7}$  \\
  & $1^{st}$ Q. (PW)\cite{suFaultTolerantQuantumSimulations2021} & (54, 3375) & 1298& $\bf 2.94 \cdot 10^{8}$  & $4.26 \cdot 10^{6}$  \\
\hline
UEG-114 & This work (min-Qu) & (114, 4096) & \bf 1535& $7.11 \cdot 10^{9}$  & $\bf 6.15 \cdot 10^{6}$  \\
  & This work (min-T) &   & 2971& $1.33 \cdot 10^{9}$  & $1.1 \cdot 10^{7}$  \\
  & $2^{nd}$ Q. (DPW)\cite{babbushEncodingElectronicSpectra2018} &   & 8314& $6.91 \cdot 10^{11}$  & $4.63 \cdot 10^{7}$  \\
  & $1^{st}$ Q. (PW)\cite{suFaultTolerantQuantumSimulations2021} & (114, 3375) & 2023& $\bf 7.83 \cdot 10^{8}$  & $7.55 \cdot 10^{6}$  \\
\hline

    \end{tabularx}
    \caption{{\bf Resource estimates for uniform electron gas systems with 14, 54, and 114 electrons and Wigner-Seitz radius of 5.} Resources are shown for the method developed in this work for two cases: one minimizes the number logical qubits (min-Qu) and the second case when one minimizes the number of Toffoli gates (min-T).  For comparison we also present resources obtained for algorithms in second quantization with dual plane waves \cite{babbushEncodingElectronicSpectra2018} and first quantization with plane waves but without the data loading~\cite{suFaultTolerantQuantumSimulations2021}. Bold font indicates smaller value.}
    \label{tab:UEG_costing}
\end{table}
\begin{table}[H]
    \centering
    \begin{tabularx}{1.0\textwidth} { 
   >{\centering\arraybackslash}c |
   >{\centering\arraybackslash}X |
   >{\centering\arraybackslash}X | 
   >{\centering\arraybackslash}X |
   >{\centering\arraybackslash}X | 
   >{\centering\arraybackslash}X
  }
\hline
  System & Method & (Electrons, Orbitals) & Logical qubits & Toffolis &  Physical qubits\\
\hline
Diamond-8 & This work (min-Qu) & (48, 4096) & \bf 759& $2.69 \cdot 10^{11}$  & $\bf 3.76 \cdot 10^{6}$  \\
  & This work (min-T) &   & 2285& $\bf 3.88 \cdot 10^{10}$  & $1.03 \cdot 10^{7}$  \\
  & $2^{nd}$ Q. (DPW)\cite{babbushEncodingElectronicSpectra2018}  &   & 8338& $1.78 \cdot 10^{14}$  & $6.27 \cdot 10^{7}$  \\
  & $1^{st}$ Q. (PW)\cite{suFaultTolerantQuantumSimulations2021} & (48, 3375) & 1508& $5.35 \cdot 10^{10}$  & $6.87 \cdot 10^{6}$  \\
\hline
Diamond-64 & This work (min-Qu) & (384, 32768) & \bf 5976& $1.51 \cdot 10^{14}$  & $4.51 \cdot 10^{7}$  \\
  & This work (min-T) &   & 21590& $3.35 \cdot 10^{12}$  & $1.33 \cdot 10^{8}$  \\
  & $2^{nd}$ Q. (DPW)\cite{babbushEncodingElectronicSpectra2018}  &   & 65702& $3.86 \cdot 10^{15}$  & $6.09 \cdot 10^{8}$  \\
  & $1^{st}$ Q. (PW)\cite{suFaultTolerantQuantumSimulations2021} & (384, 29791) & 6948& $\bf 2.19 \cdot 10^{12}$  & $\bf 4.08 \cdot 10^{7}$  \\
\hline
NiO-8 & This work (min-Qu) & (144, 4096) & \bf 1919& $1.99 \cdot 10^{12}$  & $\bf 1.03 \cdot 10^{7}$  \\
  & This work (min-T) &   & 3475& $\bf 4.09 \cdot 10^{11}$  & $1.85 \cdot 10^{7}$  \\
  & $2^{nd}$ Q. (DPW)\cite{babbushEncodingElectronicSpectra2018}  &   & 8338& $1.43 \cdot 10^{14}$  & $6.27 \cdot 10^{7}$  \\
  & $1^{st}$ Q. (PW)\cite{suFaultTolerantQuantumSimulations2021} & (144, 3375) & 2752& $7.52 \cdot 10^{11}$  & $1.47 \cdot 10^{7}$  \\
\hline
NiO-64 & This work (min-Qu) & (1152, 32768) & \bf 17505& $1.16 \cdot 10^{15}$  & $1.43 \cdot 10^{8}$  \\
  & This work (min-T) &   & 33500& $\bf 4.15 \cdot 10^{13}$  & $2.5 \cdot 10^{8}$  \\
  & $2^{nd}$ Q. (DPW)\cite{babbushEncodingElectronicSpectra2018}  &   & 65701& $3.22 \cdot 10^{15}$  & $5.84 \cdot 10^{8}$  \\
  & $1^{st}$ Q. (PW)\cite{suFaultTolerantQuantumSimulations2021} & (1152, 29791) & 18572& $4.56 \cdot 10^{13}$  & $\bf 1.39 \cdot 10^{8}$  \\
\hline 
    \end{tabularx}
    \caption{{\bf Resource estimates for diamond and nickel oxide with different number of atoms in the supercell (8 and 64 atoms).} Resources are shown for the method developed in this work for two cases: one minimizes the number logical qubits (min-Qu) and the second case when one minimizes the number of Toffoli gates (min-T). For comparison we also present resource estimates obtained for algorithms in second quantization with dual plane waves \cite{babbushEncodingElectronicSpectra2018} and first quantization with plane waves but without data loading~\cite{suFaultTolerantQuantumSimulations2021}. Bold font indicates smaller value.}
    \label{tab:materials_costing}
\end{table}

\begin{table}[H]
    \centering
    \begin{tabularx}{1.0\textwidth} { 
   >{\centering\arraybackslash}c |
   >{\centering\arraybackslash}X | 
   >{\centering\arraybackslash}X |
   >{\centering\arraybackslash}X |
   >{\centering\arraybackslash}X 
  }
\hline
Sytem &  One-norm  & One-norm & Toffoli &  Toffoli  \\ 
 & (DPW, This work)& (PW, Su \textit{et al.}~\cite{suFaultTolerantQuantumSimulations2021}) & (DPW, This work, min-T) & (PW, Su \textit{et al.}~\cite{suFaultTolerantQuantumSimulations2021}) \\ \hline
UEG-14
& {\bf 153} 
& 171 
& ${\bf 1.23 \cdot 10^{3}}$
& $1.89 \cdot 10^{3}$ 
\\ \hline
UEG-54
& $4.82 \cdot 10^{3}$ 
& ${\bf 3.42 \cdot 10^{3}} $
& $7.1 \cdot 10^{3}$ 
& ${\bf 4.73 \cdot 10^{3}} $ 
\\ \hline
UEG-114
& $1.64 \cdot 10^{4}$ 
& ${\bf 1.15 \cdot 10^{4}}$
& $9.42 \cdot 10^{3}$ 
& ${\bf 7.9 \cdot 10^{3}}$
\\ \hline
Diamond-8
& ${\bf 2.25 \cdot 10^{4}}$
& $3.87 \cdot 10^{4}$ 
& $7.02 \cdot 10^{3}$ 
& ${\bf 5.64 \cdot 10^{3}}$
\\ \hline
Diamond-64
& ${\bf 1.88 \cdot 10^{6}}$
& $2.43 \cdot 10^{6}$
& $5.81 \cdot 10^{4}$
& ${\bf 2.94 \cdot 10^{4} }$
\\ \hline
NiO-8 
& ${\bf 1.54 \cdot 10^{5}}$
& $2.76 \cdot 10^{5}$ 
& ${\bf 1.08 \cdot 10^{4}}$
& $1.11 \cdot 10^{4}$ 
\\ \hline
NiO-64
& ${\bf 1.42 \cdot 10^{7} }$
& $1.85 \cdot 10^{7}$ 
& $9.51 \cdot 10^{4}$ 
& ${\bf 8.03 \cdot 10^{4} }$
\\ \hline
    \end{tabularx}
    \caption{\textbf{One-norm and the number of Toffoli gates per block encoding.} Resources and one-norm are shown for a method developed in this work with dual plane waves and the plane-wave algorithm of Ref.~\cite{suFaultTolerantQuantumSimulations2021}. min-T means QROAM was used that minimizes the number of Toffoli gates. The number of electrons and orbitals are the same as in~\cref{tab:UEG_costing,tab:materials_costing} and for UEG-14 we used 512 basis functions. Bold font indicates smaller value.}
    \label{tab:block_encoding_break_down}
\end{table}

\section{Discussion}

\label{sec:discussion and outlook}
In this work, we have presented a Pauli LCU decomposition of the Hamiltonian in first quantization, for arbitrary basis sets. This simple form of the LCU allow us to construct an efficient block encoding which can be used with the sparse qubitization technique. In addition to the unary iteration over the electron number, our SELECT circuit involves only CCX and CCZ gates for selecting Pauli strings and the number of such gates scales only logarithmically in the size of the basis set. We observe that our approach achieves asymptotic polynomial speedup in the number of Toffoli gates compared to second quantization with the same basis set. When QROAM is implemented in such a way as to minimize the number of qubits then we see that already for an active space of 16 electrons and 32 orbitals of the \fes molecule, the first quantization algorithm requires fewer qubits compared to its second quantization counterpart. We observe these asymptotic improvements because the one-norm of the Pauli LCU decomposition scales better, but the large prefactor due to explicit dependence on the number of electrons squared can only be overcome for large basis sets.

In recent years, several efficient LCU decompositions of the Hamiltonian in second quantization and corresponding block-encodings have been developed~\cite{berryQubitizationArbitraryBasis2019a,vonburgQuantumComputingEnhanced2021a,leeEvenMoreEfficient2021a, ivanovQuantumComputationPeriodic2023,rubin_fault-tolerant_2023}. Exactly the same methods, such as factorization of the Hamiltonian's matrix elements, can of course be applied in first quantization. It is of great importance to understand if the first quantization approach can bring advantage beyond asymptotic improvements over second quantization when such factorizations are applied.

We have also applied our method to the DPW basis set, which is more suitable for periodic systems. In that case, the SELECT and PREP circuits can be slightly optimized compared to the generic quantum chemistry case. The resource estimates for realistic materials show that in some cases we outperform in some metrics (either qubit or Toffoli counts) the previous first quantization approach of Refs.~\cite{babbushQuantumSimulationChemistry2019,suFaultTolerantQuantumSimulations2021}, which used regular PW. Our approach is more efficient in Toffoli and qubit count than the second quantization approach with DPW~\cite{babbushEncodingElectronicSpectra2018} on all considered examples. However, we recognise that there are regimes when second quantization proves advantageous, for example, when one considers systems with the same number of orbitals as the number of electrons (half-filling).

Our approach with DPW employs classical data loading, yet it still provides quantum resources comparable to the approach of Refs.~\cite{babbushQuantumSimulationChemistry2019,suFaultTolerantQuantumSimulations2021} in the physically interesting regimes. The reasons for this are the sub-linear scaling of the one-norm of the Pauli LCU in the basis set size and the diagonal form of the Coulomb interaction which significantly reduces the amount of information needed to specify the Hamiltonian. For example, for NiO we achieve a lower T-gate count and the number of logical qubits is only a factor of 1.8 larger than that obtained with pure PW approach, for a PW cutoff that is suitable for use with pseudopotentials. However, since we used data loading, our approach can be used with norm-conserving pseudopotentials, and projector augmented-wave method (or unitary version of PAW~\cite{ivanov_upaw_2024}~\footnote{Conventional PAW approach~\cite{blochl_projector_1994} introduces the non-orthogonal transfromations which do not allow for a simple Hamiltonian expression using plane wave or dual plane waves in first quantization~\cref{eq:full_ftq_chem_hamiltonian_with_spin} and as a result PAW introduces Hamiltonian with non-orthogonal eigenstates. However, the unitary version of PAW (UPAW)~\cite{ivanov_upaw_2024} overcomes these issues.}), because the only change would be in the calculation of matrix elements on a classical computer. 
As is shown in several density functional theory calculations~\cite{delta_dft,lejaeghere_reproducibility_2016,prandini2018precision,bosoni_how_2023}, PAW approach demonstrates smaller approximation error than norm-conserving pseudopotentials and often requires fewer number of PW in order to reach the convergence.
While there has been generalization of Refs.~\cite{babbushQuantumSimulationChemistry2019,suFaultTolerantQuantumSimulations2021} on norm-conserving Goedecker-Teter-Hutter pseudopotentials~\cite{goedecker_separable_1996,hartwigsen_relativistic_1998}, it is not clear if such an approach can be easily generalized for PAW method or even other pseudopotentials without the data loading.

We also speculate that the algorithms presented here can be further improved. For example, in the DPW basis set, there are only $3D$ unique Hamiltonian matrix elements and all other matrix elements could be restored by using the translational symmetry. We have not considered this here. One possible way to take this into account is to load unique original matrix elements from QROAM and then calculate the Pauli LCU coefficients (\cref{eq:dpw_plcu_coeff_one_body,eq:dpw_plcu_coeff_two_body}) on the quantum computer (quantum Pauli decomposition). This would reduce the data loading which in turn would make Toffoli cost of the PREPARE to scale as $O(D^{0.5})$.

The Pauli LCU decomposition of the Hamiltonian~\eqref{eq:canon_lcu_h1},~\eqref{eq:canon_lcu_h2} is generic and it can be applied equally to either bosonic or fermionic or even mixed systems. This in turn opens up for possibilities to apply the approach developed here to other application areas such as studying vibrational properties of molecules and materials~\cite{mcardleDigitalQuantumSimulation2019, sawayaResourceefficientDigitalQuantum2020, sawayaLongtermQuantumAlgorithmic2021, trenevRefiningResourceEstimation2023}. We also hope that the explicit form of the Pauli LCU decomposition can be used to reduce resources in other quantum algorithms such as Trotterization~\cite{suzuki1991general, childs2018toward} or, more generally, Hamiltonian simulation~\cite{abramsSimulationManyBodyFermi1997a}.

\section{Code Availability}
The PySCF~\cite{sun_recent_2020, sun_pyscf_2018, sun_libcint_2015}, Atomic Simulation Environment (ASE)~\cite{ase_new,ase_old},  GPAW~\cite{mortensen_gpaw_an_open_2024, Enkovaara2010,Mortensen2005}, OpenFermion~\cite{McClean_2020} and VMD~\cite{humphrey_vmd_1996} software were used to produce the results for this paper. PySCF, ASE, GPAW and OpenFermion are open source software.

\section{Data Availability}
The electron integrals in molecular orbital basis sets and dual plane waves used to produce results for this paper are available at Zenodo repository[https://doi.org/10.5281/zenodo.14906508]. 

\section{Acknowledgements}

We thank Nick Blunt for discussions, carefully reading the manuscript and providing valuable suggestions, as well as Andrew Patterson for assistance with quantum error correction resource estimates. We also thank Matt Ord for improvements to the resource estimation software and assistance deriving the LCU decomposition.
The work presented in this paper was part funded by a grant from Innovate UK under the ’Feasibility Studies in Quantum
Computing Applications’ competition (Project Number 10074148).
M.B. is a Sustaining Innovation Postdoctoral Research Associate at Astex Pharmaceuticals and thanks Astex Pharmaceuticals for funding, as well as his Astex colleague Patrick Schoepf for his support.

\section{Author Contributions}
 T.N.G. derived LCU decomposition with input from A.V.I. and B.K.B., and wrote~\cref{sec:lcu,apx:derivation_lcu_twobody}. T.N.G. and A.V.I. carried out quantum chemistry calculations with input from R.I.. T.N.G carried out numerical analysis for~\cref{subsec:scaling_properties_mol,subsec:quantum_resources_mol} with input from A.V.I. and R.I.. M.B. and A.V.I. carried out numerics for~\cref{subsec:scaling_properties_dpw,subsec:resources_dpw}. A.V.I. and C.S. worked out the SELECT operator, C.S. derived the detailed cost of the algorithm and wrote~\cref{subsec:arbitrarybasisset_blockencoding,apx:amplification}. M.B. contributed to the costing of amplitude amplification (~\cref{apx:amplification}). R.I. wrote about the iron-sulphur complex. A.V.I. wrote the final version of the paper with input from all authors. A.V.I. conceived of and supervised the project.

\section{Methods}
\label{sec:appendix}
\subsection{Two-Body LCU Derivation}
\label{apx:derivation_lcu_twobody}
In this Section, we derive \cref{eq:twobody_lcu_with_spin}, starting from the second term on the right-hand side of \cref{eq:full_ftq_chem_hamiltonian_with_spin}. First, we express the tensor product of operators in terms of $X$ and $Z$ Pauli operators in the same way as Ref.~\cite{paulidecompriverlane2024}:
\begin{equation}
    \left(\ketbra{p}{q})_i \right(\ketbra{r}{s})_j =
    \left(\sum_{u=0}^{D-1} \frac{1}{D} \prod_{k=0}^{M-1}(-1)^{q_k\wedge u_k}X_{iM+k}^{p_k\oplus q_k}Z_{iM+k}^{u_k}\right) 
    \left(\sum_{v=0}^{D-1} \frac{1}{D} \prod_{l=0}^{M-1}(-1)^{s_l\wedge v_l}X_{jM+l}^{r_l\oplus s_l}Z_{jM+l}^{v_l}\right).
\end{equation}
\noindent We have neglected the spin indices, because these cancel to the identity the same as in the one-body term. Next, we take out the factors of $-1$, which are the matrix elements of Hadamard matrices (see definition~\cref{eq:hadamard}),
\begin{equation}\label{eq:pqrs_expansion}
    \left(\ketbra{p}{q}\right)_i\left(\ketbra{r}{s}\right)_j =
    \sum_{u=0}^{D-1} \sum_{v=0}^{D-1} \frac{1}{D^2} \big(H^{\otimes M}\big)_{qu} \big(H^{\otimes M}\big)_{sv}
    \left(\prod_{k=0}^{M-1}X_{iM+k}^{p_k\oplus q_k}Z_{iM+k}^{u_k}\right)
    \left(\prod_{l=0}^{M-1}X_{jM+l}^{r_l\oplus s_l}Z_{jM+l}^{v_l}\right).
\end{equation}
Substituting \cref{eq:pqrs_expansion} into the two-body term of \cref{eq:full_ftq_chem_hamiltonian_with_spin}, we find
\begin{equation}
    \hat{H}_{\text{two-body}}= \frac{1}{2}
    \sum_{p,q,r,s=0}^{D-1}\sum_{u,v=0}^{D-1}
    \frac{1}{D^2}  h_{pqrs} \big(H^{\otimes M}\big)_{qu} \big(H^{\otimes M}\big)_{sv}\sum_{i\ne j}^{N-1}
    \left(\prod_{k=0}^{M-1}X_{iM+k}^{p_k\oplus q_k}Z_{iM+k}^{u_k}\right)
    \left(\prod_{l=0}^{M-1}X_{jM+l}^{r_l\oplus s_l}Z_{jM+l}^{v_l}\right).
\end{equation}
\noindent Next, we change sum index from $p$ to $g=p\oplus q$ and from $r$ to $f=r\oplus s$, giving
\begin{equation}
    \hat{H}_{\text{two-body}}=
    \sum_{u,v,g,f=0}^{D-1}\omega_{gufv}
    \frac{1}{2}\sum_{i\ne j}^{N-1}\left(\prod_{k=0}^{M-1}X_{iM+k}^{g_k}Z_{iM+k}^{u_k}\right)
    \left(\prod_{l=0}^{M-1}X_{jM+l}^{f_l}Z_{jM+l}^{v_l}\right),
\end{equation}
\noindent where
\begin{equation}
    \omega_{gufv}=\frac{1}{D^2}\sum_{q,s=0}^{D-1}
     h_{g\oplus q,q,f\oplus s,s} \big(H^{\otimes M}\big)_{qu} \big(H^{\otimes M}\big)_{sv}.
\end{equation}
\noindent Since $i\ne j$, the two Pauli strings always commute, and as a result
\begin{equation}
    \hat{H}_{\text{two-body}}=
    \sum_{u,v,g,f=0}^{D-1}\omega_{gufv}
    \frac{1}{2}\sum_{i\ne j}^{N-1}
    \prod_{k=0}^{M-1}X_{Mi+k}^{g_k}Z_{Mi+k}^{u_k}
    X_{Mj+k}^{f_k}Z_{Mj+k}^{v_k}.
\end{equation}
\noindent This is the result of \cref{eq:twobody_lcu_with_spin} with different labels for the indices: $u,v,g,f$ are $p,q,r,s$, respectively.
To turn the two-body LCU coefficients calculation into a matrix multiplication, we form a $D^2\times D^2$ Hadamard matrix:
\begin{equation}
    \left(\prod_{k=0}^{M-1}(-1)^{q_k\wedge u_k}\right)
    \left(\prod_{l=0}^{M-1}(-1)^{s_l\wedge v_l}\right)
    =\big(H^{\otimes M}\big)_{qu}\big(H^{\otimes M}\big)_{sv}=\big(H^{\otimes 2M}\big)_{Dq+s, Du+v}.
\end{equation}

\noindent If we rearrange the matrix elements with 

\begin{equation}
    \Tilde{h}_{Dg+f, Dq+s}=h_{g\oplus q,q,f\oplus s,s}, 
\end{equation}

\noindent the LCU coefficients can be found with matrix multiplication,

\begin{equation}\label{eq:twobody_mels_matmul}
    \omega_{gufv}=\frac{1}{D^2}\sum_{q,s=0}^{D-1}
    \Tilde{h}_{Dg+f, Dq+s} \big(H^{\otimes 2M}\big)_{Dq+s, Du+v}
\end{equation}

\noindent Using the fast Hadamard transform, calculating these matrix elements scales as $\oo(D^4\log D)$, rather than $\oo(D^6)$ for basic matrix multiplication.

\subsection{Amplitude Amplification for equal superpositions}

\label{apx:amplification}

The equal superposition state
\begin{equation}
    \frac{1}{\sqrt{L}}\sum_{l=0}^{L-1} \ket{l}
\end{equation}
needed in PREP can be prepared trivially with Hadamards if $L$ is a power of 2. Otherwise, it is prepared using amplitude amplification; the procedure for this state is described in \cite{leeEvenMoreEfficient2021a}.

In this appendix, we will adapt that procedure and cost the preparation of the equal superposition state
\begin{equation}
    \frac{1}{\sqrt{N(N-1)}}\sum_{i\neq j}^{N-1}\ket{i}\!\ket{j}
    \label{eq:ij superpos}
\end{equation}
also required in PREP. We use a single iteration of Grover search \cite{Nielsen_Chuang_2010} 
along the lines of Fig.~3 from \cite{leeEvenMoreEfficient2021a}.

In order to achieve a success probability close to one, we augment \eqref{eq:ij superpos} by an ancilla rotated by a classical tunable angle $\beta$, such that the desired state is
\begin{equation}
    \ket{\phi} = \frac{1}{\sqrt{N(N-1)}}\sum_{i\neq j}^{N-1}\ket{i}\!\ket{j} \otimes R_Y^\dag(\beta)\ket{0}
\end{equation}
The starting point for Grover search of this state (including the rotated ancilla) is an equal superposition over all computational basis states
\begin{equation}
    \ket{\psi} = \frac{1}{\sqrt{2^{\lceil\log_2 N\rceil}2^{\lceil\log_2 N\rceil}}}\sum_{i,j=0}^{2^{\lceil\log_2 N\rceil}-1}\ket{i}\ket{j} \otimes \ket{+}.
\end{equation}

The circuit we use to perform the Grover iteration is the following:
\begin{equation} \label{equalSuperpositionCircuit}
\begin{tikzpicture}
\begin{yquant}
qubit {$\ket{0}^{\bigotimes \lceil\log_2 N\rceil}$} a;
qubit {$\ket{0}^{\bigotimes \lceil\log_2 N\rceil}$} b;
qubit {$\ket{0}$} r;

qubit {$\ket{0}$} anca;
qubit {$\ket{0}$} ancb;
qubit {$\ket{0}$} anceq;

qubit {succ\ $\ket{0}$} succ;

["north:$\lceil\log_2 N\rceil$" {font=\protect\footnotesize, inner sep=0pt}]
slash a,b;

h a,b,r;

text {$i$} a;
text {$j$} b;

[multictrl, name=oraclecompute]
box {$\oplus($\Ifcase\idx\relax$i<N$\Or$j<N$\Or$i=j$\Fi)} anca, ancb, anceq ~ a,b;

box {$R(\beta)$} r;
phase {} anca | ancb ~ r, anceq;
box {$R^\dag(\beta)$} r;

[multictrl, name=oracleuncompute]
box {$\oplus($\Ifcase\idx\relax$i<N$\Or$j<N$\Or$i=j$\Fi)} anca, ancb, anceq ~ a,b;

\node[draw=red, dashed, fit=(oraclecompute-2) (oraclecompute-n0) (oraclecompute-1) (oracleuncompute-2) (oracleuncompute-n0) (oracleuncompute-1), "$(I-2\ket{\phi}\!\bra{\phi})$"] {};

hspace {2mm} -;

[name=groverref1]
H a,b,r;
[operator style={/yquant/every negative control}]
phase {} a ~ b,r;
[name=groverref2]
H a,b,r;

\node[draw=red, dashed, fit=(groverref1-0) (groverref1-2) (groverref2-0) (groverref2-2), "$(I-2\ket{\psi}\!\bra{\psi})$"] {};


[multictrl, name=testuncompute]
box {$\oplus($\Ifcase\idx\relax$i<N$\Or$j<N$\Or$i=j$\Fi)} anca, ancb, anceq ~ a,b;

cnot succ | anca, ancb ~ anceq;

[multictrl, name=testuncompute]
box {$\oplus($\Ifcase\idx\relax$i<N$\Or$j<N$\Or$i=j$\Fi)} anca, ancb, anceq ~ a,b;
\end{yquant}
\end{tikzpicture}
\end{equation}
It closely follows the textbook circuit \cite{Nielsen_Chuang_2010}. Initially, the Hadamards prepare the  state $\ket{\psi}$. This is followed by a Grover iteration, consisting of reflections about the target state, $(I-2\ket{\phi}\!\bra{\phi})$, and about the initial state, $(I-2\ket{\psi}\!\bra{\psi})$. Finally, the desired state is flagged in the success qubit.

The overlap of desired state and initial state is
\begin{equation}
    \cos\alpha := \langle\phi\vert\psi\rangle = \sqrt{\frac{N(N-1)}{2^{\lceil\log_2 N\rceil}2^{\lceil\log_2 N\rceil}}}
\sin (\beta/2+\pi/2).
\end{equation}
The output of a single iteration Grover circuit is a state \cite{Nielsen_Chuang_2010}
\begin{equation}
    \cos (3\alpha) \ket{\phi} + \sin (3\alpha) \ket{\phi^\perp},
\end{equation}
where $\ket{\phi^\perp}$ is some state orthogonal to $\ket{\phi}$; the success probability has been amplified from $\cos\alpha$ to $\cos(3\alpha)$.
Thus, one can ensure a high success probability by choosing $\beta$ such that $\cos(3\alpha) = 1 \Rightarrow \cos\alpha=-1/2$.
This is always possible because
\begin{equation}
\sqrt{\frac{N(N-1)}{2^{\lceil\log_2 N\rceil}2^{\lceil\log_2 N\rceil}}} > \frac{1}{2} \, , 
\end{equation}
so it can always be scaled down by $\sin(\beta/2+\pi/2)$ to reach $\cos\alpha=-1/2$.

The optimal value of $\beta$ must be approximated to accuracy $\epsilon$ when compiling the rotations $R_Y(\beta)$. In principle, they could be compiled directly into a Clifford+$T$ gate set \cite{ross2016optimal} using $4\log_2(1/\epsilon) + O(\log \log(1/\epsilon))$ $T$-gates. However, \cite{leeEvenMoreEfficient2021a} choose the approach of \cite{sandersCompilationFaultTolerantQuantum2020} Appendix A, adding a binary representation of $\beta$ into a phase-gradient ancilla register, which has a similar $T$ count. Yet, the latter reference mentions that for a classically given $\beta$, like here, compilation into Clifford+$T$ is slightly faster. For consistency with literature, we follow the approach of \cite{leeEvenMoreEfficient2021a} using the phase gradient register.

The cost of this circuit is as follows: The Hadamard gates are Clifford gates and therefore require no Toffolis. The two inequality tests $(i<N)$ and $(j<N)$ are implemented as described in appendix H of \cite{berryQubitizationArbitraryBasis2019a}, where we can save one Toffoli because $N$ is constant, for a Toffoli cost of $2(\lceil \log_2 N \rceil -1)$. The final equality test requires $\lceil \log_2 N \rceil$ Toffolis. If $\eta_N$ is the largest power of 2 that divides $N$, the cost of each $i<N$ inequality test is reduced by $\eta_N$ Toffolis (by ignoring the lowest $\log_2\eta_N$ bits/qubits of $N$ and $\ket{i}$).  Each test also requires $\lceil \log_2 N \rceil -1$ ancilla not depicted in (\ref{equalSuperpositionCircuit}), and can be uncomputed without additional Toffoli cost using the out-of-place adders from \cite{gidney2018}. (Essentially, the first $\oplus(i<N)$ is only the computation step of the addition. The ancillas are kept alive during the CCCZ gate, and the second $\oplus(i<N)$ is the uncomputation step of the addition). The total cost of all the (in)equality tests in the circuit (\ref{equalSuperpositionCircuit}) is therefore $6 \lceil \log_2 N \rceil-4 \eta_N-4$ Toffolis and $3\lceil \log_2 N \rceil-3$ ancilla.

The rotation by $\beta$ using $b_N$ bits of precision (typically $b_N = 8$) costs $b_N -3 $ Toffolis, since the angle is known classically, plus a one-time cost for preparing a catalytic state that can be reused for rotations throughout the circuit and is thus negligible. It also requires $b_N$ ancilla.
The reflection around $\ket{\phi}$ requires 2 Toffolis, the reflection around $\ket{\psi}$ needs $2 \lceil \log_2 N \rceil-1$ Toffolis, and the flagging of the success qubit requires another 2 Toffolis. The total Toffoli cost of the equal superposition circuit is therefore
\begin{align} 
    8 \lceil \log_2 N \rceil - 4\eta_N+2 b_N -7 \, 
\end{align}
Toffolis, and requires the use of $3\lceil \log_2 N \rceil + b_N -3$ ancillas in parallel. We can reuse the ancilla from the QROAM procedure for most of these ancilla, with two exceptions: We need to keep the phase-gradient register of size $b_N$, so we do not have to prepare it every time, and the success qubit and rotation qubit flag the block encoding and therefore cannot be reused either. The number of additional ancilla we need for this procedure is therefore
\begin{align}
    b_N + 2 \, .
\end{align}

\bibliography{main.bib}

\clearpage

\end{document}